\documentclass[sigconf,nonacm]{acmart}

\usepackage{xargs}      
\usepackage{soul}       
\usepackage{xcolor}     
\usepackage{enumitem}   
\usepackage{subcaption} 
\usepackage{multirow}   
\usepackage{algorithm}
\usepackage{algorithmic}
\usepackage{graphicx}
\usepackage{makecell}

\definecolor{commentgray}{RGB}{120, 120, 120}

\usepackage{listings}
\usepackage[most]{tcolorbox}
\definecolor{keycolor}{RGB}{30, 100, 200}
\definecolor{stringcolor}{RGB}{50, 150, 50}
\definecolor{backcolor}{RGB}{248, 248, 248}
\lstdefinelanguage{json}{
    basicstyle=\footnotesize\ttfamily,
    stepnumber=1,
    numbersep=4pt,
    showstringspaces=false,
    breaklines=true,
    frame=none,
    literate=
     *{0}{{{\color{blue}0}}}{1}
      {1}{{{\color{blue}1}}}{1}
      {2}{{{\color{blue}2}}}{1}
      {3}{{{\color{blue}3}}}{1}
      {4}{{{\color{blue}4}}}{1}
      {5}{{{\color{blue}5}}}{1}
      {6}{{{\color{blue}6}}}{1}
      {7}{{{\color{blue}7}}}{1}
      {8}{{{\color{blue}8}}}{1}
      {9}{{{\color{blue}9}}}{1}
      {:}{{{\color{black}:}}}{1}
      {,}{{{\color{black},}}}{1}
      {\{}{{{\color{orange}\{}}}{1}
      {\}}{{{\color{orange}\}}}}{1}
      {[}{{{\color{orange}[}}}{1}
      {]}{{{\color{orange}]}}}{1},
}
\definecolor{darkgreen}{rgb}{0.0, 0.5, 0.0}  
\lstdefinelanguage{markdown}{
    basicstyle=\tiny\ttfamily,
    morekeywords={\#, \*\*, \*, -}, 
    keywordstyle=\color{blue},
    sensitive=false,
    morecomment=[l]{>}, 
    commentstyle=\color{gray}\itshape,
    morestring=[b]"
}

\newcommand{\crowdgenui}{\textsc{AlignUI}}

\AtBeginDocument{%
  }

\begin{document}

\title{\crowdgenui{}: A Method for Designing LLM-Generated UIs Aligned with User Preferences}

\author{Yimeng Liu}
\authornote{This work was done during an internship at Adobe Research.}
\affiliation{%
  \institution{University of California, Santa Barbara}
  \city{Santa Barbara}
  \country{USA}}
\email{yimengliu@ucsb.edu}

\author{Misha Sra}
\affiliation{%
  \institution{University of California, Santa Barbara}
  \city{Santa Barbara}
  \country{USA}}
\email{sra@ucsb.edu}

\author{Chang Xiao}
\affiliation{%
  \institution{Adobe Research}
  \city{San Jose}
  \country{USA}}
\email{cxiao@adobe.com}

\renewcommand{\shortauthors}{Liu et al.}


\begin{abstract}
Designing user interfaces that align with user preferences is a time-consuming process, which requires iterative cycles of prototyping, user testing, and refinement. Recent advancements in LLM-based UI generation have enabled efficient UI generation to assist the UI design process. We introduce \crowdgenui{}, a method that aligns LLM-generated UIs with user tasks and preferences by using a user preference dataset to guide the LLM's reasoning process. The dataset was crowdsourced from 50 general users (the target users of generated UIs) and contained 720 UI control preferences on eight image-editing tasks. We evaluated \crowdgenui{} by generating UIs for six unseen tasks and conducting a user study with 72 additional general users. The results showed that the generated UIs closely align with multiple dimensions of user preferences. We conclude by discussing the applicability of our method to support user-aligned UI design for multiple task domains and user groups, as well as personalized user needs. 
\end{abstract}



\begin{teaserfigure}
    \centering
    \includegraphics[width=\linewidth]{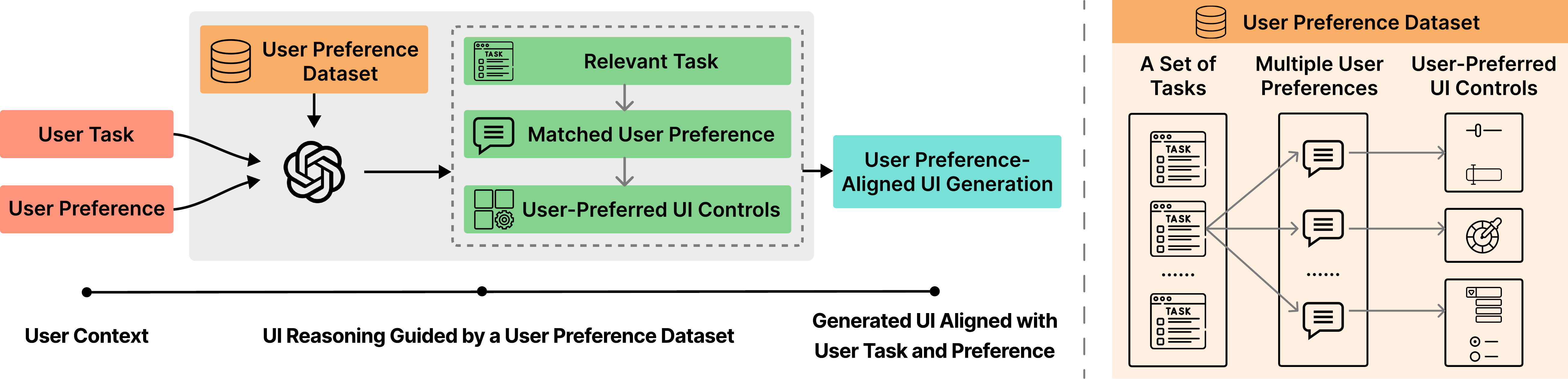}
    \caption{\crowdgenui{} supports user preference-aligned UI design with LLM-based UI generation (left). With the user context, which contains user task and preference, an LLM is prompted to reason UIs guided by a user preference dataset (right). This dataset consists of a set of tasks and multiple user preferences for each task. For each user preference, the dataset contains user-preferred UI controls. We collect this dataset by crowdsourcing (\S~\ref{sec:crowdsourcing}). With the dataset, the UI reasoning follows a multi-stage approach, where the LLM first identifies tasks in the dataset that are similar to the user task, then finds user preference that matches the user-specified preference, and finally reasons and generates UI controls that align with the user task and preference.}
    \Description{This figure shows a flowchart of user preference-aligned UI generation guided by a user preference dataset on the left, and a illustration of the user preference dataset on the right. The flowchart takes user context as input, uses it to reason UIs guided by the user preference dataset, and generates UIs aligned with the user context (user task and preference). The user preference dataset contains a set of tasks, multiple user preferences, and user-preferred UI controls.}
    \label{fig:teaser}
\end{teaserfigure}

\maketitle


\section{Introduction} \label{sec:introduction}
Designing and creating user interfaces (UIs) that resonate with their target users' preferences is a cornerstone of modern digital interaction. The UI design process with user preference taken into account is labor-intensive, which requires significant cross-disciplinary expertise to bridge the silos between creative vision, technical feasibility, and user alignment~\cite{mcinerney2000ui}. Designers need to hand-craft the design of each UI control and make sure it aligns with the intended functionality~\cite{lu2022bridging}. Beyond these functional requirements of UI interactions, designers also need to navigate tradeoffs between aesthetic appeal and perceived usability, as visual presentation has been shown to fundamentally impact user performance and satisfaction~\cite{tractinsky2000beautiful}. Aligning these designs with target users' needs and preferences further complicates this process, which often requires iterative cycles of prototyping, user testing, and refinement~\cite{shneiderman2010designing, giacomin2014human}.

To assist the UI design process, tools have been developed to support UI programming~\cite{myers1995user}, digital UI design and prototyping~\cite{photoshop, figma}, and AI-based automatic UI generation~\cite{li2019layoutgan, huang2019swire}. However, these tools either require extensive manual efforts to craft and iterate on UI designs, or delegate the UI design process entirely to AI models, which risks the agency of UI designers. To balance automation and human agency, recent research has explored human-AI collaboration in the UI design and prototyping process~\cite{subramonyam2021protoai}. Though helpful to bridge human and AI capabilities and allow human-centered UI design experience, the AI-generated UI design suggestions do not always align with designers' goals or target users' preferences, such that the designers often need to refine AI outputs through long-term, back-and-forth interaction.

Recent advancements in large language model (LLM)-based UI generation have shown great potential to simplify the UI design process. UI designers can generate functional UIs from simple natural language prompts~\cite{leviathangenerative}. Such capability can enable cheaper UI prototyping and iteration with fast UI generation and preliminary user insights used to train the LLMs. These LLM-generated UIs have been explored in desktop applications~\cite{liu2025auggen} and web pages~\cite{leviathangenerative}, as well as to support UI prototyping~\cite{lu2025misty}, UI design space search~\cite{park2025leveraging}, and user experience (UX) practices~\cite{chen2025genui}. While promising, the UIs generated by these models are often generic. They frequently fail to capture the nuanced, subjective user preferences, leading to interfaces that may be technically functional but experientially disconnected with target users (based on our formative study in Section~\ref{sec:prelim_study}). Even with user instructions to refine the LLM-generated outputs, LLMs can still struggle to effectively follow these instructions to produce user-desired results~\cite{zhou2023instruction}. In addition, LLM-generated content is inconsistent, which can be different from run to run even when the temperature parameter is set as zero~\cite{yuan2025understanding}, causing the generated UIs difficult to predict~\cite{vaithilingam2024dynavis}. The generic, inconsistent LLM-generated UIs go contrary to the desired benefits of automation for user preference-aligned UI generation since designers still need extensive editing to refine the LLM-generated prototypes and align these prototypes with target users. 

To bridge this gap, prior research typically relies on reinforcement learning from human feedback (RLHF)~\cite{ouyang2022training} with direct preference optimization (DPO)~\cite{rafailov2023direct} or group relative policy optimization (GRPO)~\cite{shao2024deepseekmath} to align LLM outputs with human preference and intent. However, these techniques are often applied in the LLM post-training stage and known to be data intensive and computationally expensive~\cite{kumar2025llm}. This creates a high barrier to entry and limits the ability of UI designers to rapidly adapt models to evolving user preferences or needs.

To address these challenges, \textbf{we introduce \crowdgenui{}, a lightweight UI generation approach that aligns LLM-generated UIs with UI users' preference to support the UI design process}. Our approach operates in the LLM inference stage without the need for expensive model post-training. Motivated by the effective use of retrieval-augmented generation (RAG) to reduce LLM hallucination through an external knowledge base~\cite{lewis2020retrieval}, our approach injects a user preference dataset directly into the LLM's reasoning process to guide UI generation. Specifically, this dataset is organized as three layers (Figure~\ref{fig:teaser} right): the \emph{task} (the goal to achieve and specific task requirements), \emph{user preference aspects} (the aspects that the user prefers to complete the task), and \emph{user-preferred UI controls} (the UI controls that the user prefers regarding each user preference aspect). With this dataset, the LLM's reasoning is guided by a multi-stage approach to generate UI candidates that align with both user task and preference by identifying similar tasks in the dataset, finding matched user preference, reasoning user preference-aligned UI controls, and generating the code to implement them (Figure~\ref{fig:teaser} left). 

To evaluate our approach, we used it to generate UIs for image editing tasks, which require instant visual feedback and iterative UI interaction to achieve user-desired outcomes. The user preference dataset to guide UI generation was crowdsourced from 50 general users (the target users of generated UIs) on eight tasks. The user preferences covered three user preference aspects that were derived from established UI design principles~\cite{lidwell2010universal, mckay2013ui}, namely, \emph{predictability} (achieve results without unexpected outcomes), \emph{efficiency} (perform tasks with minimal time and effort), and \emph{explorability} (explore multiple possibilities to complete the task). With our generated UIs, we conducted a user study with 78 additional participants on six new tasks. The study results showed that UIs generated using our approach were rated as more aligned with UI users' preferences compared to those generated by the LLM alone. These results highlight the effectiveness of our approach to advance user preference-aligned UI generation. In summary, our main contributions are as follows:

\begin{itemize}[left=0pt]
    \item A \textbf{formative study} that surfaces the user preference alignment gaps in LLM-generated UIs and commercial software UIs. 
    \item A lightweight, \textbf{user preference-aligned UI generation method}, which is guided by a user preference dataset during LLM reasoning, to address the identified user preference alignment gaps.
    \item A \textbf{crowdsourced user preference dataset} that captures 720 pieces of user preferences from 50 general users, who are the target users of generated UIs.
    \item \textbf{Generated UIs} using our UI generation method, and a \textbf{user study} with 72 additional general users to evaluate the generated UIs, which demonstrates the effectiveness of our UI generation method to align with user preferences.
\end{itemize}

\section{Related Work} \label{sec:relatedwork}
Our work is grounded in UI design process, tools and techniques to support the UI design process, LLM-based UI generation, and LLM alignment with humans. 

\textbf{UI design process.}
UI design is a long-term process that requires designers' knowledge and expertise to optimize multiple UI design aspects for the target users. In this process, designers often need to follow theoretical foundations and UI design principles, such as Shneiderman's ``Eight Golden Rules''~\cite{shneiderman2010designing} and Nielsen's usability components~\cite{nielsen1994usability} to bridge the ``Gulf of Execution and Evaluation''~\cite{norman1988psychology}. To create UIs based on their design, designers are required to hand-craft/sketch every UI control to define each UI control's functionality, to experiment with the spatial arrangement, and to optimize user interaction across screen sizes and device constraints~\cite{gould1985designing, olsson2004active}. The created UIs should achieve the intended functionality~\cite{lu2022bridging}, as well as meet aesthetic and usability requirements for their audience~\cite{tractinsky2000beautiful}. Aligning designed UIs with target users (such as target users' preferences) requires iterative cycles of prototyping, implementation, user evaluation (e.g., usability test, survey, interview), and refinement to minimize the gap between the intended design and user adoption~\cite{shneiderman2010designing, giacomin2014human}. 

\textbf{UI design support tools and techniques.}
Since the UI design process is known to be time-consuming and error-prone~\cite{myers2000past}, tools have been developed to facilitate this process. In 1990s, Myers~\cite{myers1995user} has studied various kinds of UI design tools to support productivity for UI programmers. Following the early efforts, commercial software tools started to arise to support digital UI design, such as Photoshop~\cite{photoshop}, Sketch~\cite{sketch}, and Figma~\cite{figma}. More recent work has leveraged AI techniques to allow data-centered UI design. With UI datasets, such as SWIRE~\cite{huang2019swire} and RICO~\cite{deka2017rico}, deep learning approaches have been developed to use these datasets to generate UIs~\cite{li2019layoutgan, moran2018machine}, assist UI design space search~\cite{huang2019swire}, and support UI element understanding~\cite{wang2021screen2words}. 

Nevertheless, these tools and techniques either require extensive manual efforts to sketch and implement UI elements and layout, or delegate the UI design process to AI models, which isolates this process from the designers. While recent work has explored human-AI collaboration to assist the UI design and prototyping stages to balance AI automation and human agency, such as ProtoAI~\cite{subramonyam2021protoai}, the initial AI-generated content remains generic for different users and the user's role focuses on refining these AI outputs through back-and-forth interaction. The direct incorporation of target user understanding, such as user preferences, into the AI's UI design generation process is largely missing. To fill this gap, we collect user data and directly inject it into the UI design generation process to align AI-generated UI designs with target users, and probe the user data from the user preference perspective.

\textbf{LLM-based UI generation.}
With the advancement of LLMs, recent work has explored automatically generating UIs with LLMs to assist the UI design process. This approach takes natural language prompts as input and generates UIs tailored to users' task descriptions and requirements, which is based on the LLM's reasoning and code generation capabilities. For instance, DynaVis~\cite{vaithilingam2024dynavis} generates dynamic UIs for visualization editing, Biscuit~\cite{cheng2024biscuit} generates ephemeral UIs to help users follow machine learning tutorials, AugGen~\cite{liu2025auggen} generates scaffolded UIs in professional software to support learning, Jelly~\cite{cao2025generative} infers data models from the user input prompt to guide UI generation for end users, and Leviathan et al.~\cite{leviathangenerative} explored UI generation for web pages and released an expert-crafted dataset for UI evaluation. 

Although LLMs are trained on large-scale human-produced data, which can be helpful to model target users, most prior work using standalone LLMs does not fully consider user preferences of specific user groups. This can result in generated UIs that, while functional, may not be the most effective for specific users or aligned with their preferences~\cite{lu2022bridging}. Although existing research has explored human-in-the-loop interaction design to enable UI designers and end users to adjust LLM-generated UIs by natural language instruction and direct manipulation~\cite{yuan2024towards, cao2025generative, rabi2025designing}, the fine-tuned outputs do not always align with user intent due to the nature of inconsistent LLM generation~\cite{barke2023grounded} and the challenges in LLM's instruction following~\cite{zhou2023instruction, zeng2023evaluating}. In this work, we tackle the user preference alignment challenge in LLM-generated UIs by guiding the UI generation process with a user preference dataset containing rich user preference data from target users, which reduces the user's burden on tuning LLM outputs. 

\textbf{LLM alignment with humans.}
Aligning LLM outputs with human values has been a critical research topic to ensure safety and user satisfaction of LLM-generated text, code, and UIs~\cite{wang2023aligning}. To enable LLM alignment with humans, recent research has explored techniques including reinforcement learning from human feedback~\cite{ouyang2022training} with direct preference optimization~\cite{rafailov2023direct} or group relative policy optimization~\cite{shao2024deepseekmath}. These techniques are typically used in LLM post-training to reward LLMs that produce human-aligned outputs and gradually guide the model outputs towards user-desired directions. While effective and widely used, they are known to be expensive with regard to data collection, computation resources, and training time~\cite{kumar2025llm}. Data needs to be collected and labeled with human raters and used to fine-tune LLMs through long-term training, making these techniques hard to be employed by UI designers and end users to align LLM-generated content with their specific needs. 

Motivated by retrieval-augmented generation (RAG), which has been used to mitigate the hallucination problem by offering LLMs an external knowledge base~\cite{lewis2020retrieval}, our work provides the LLM with a user dataset, which focuses on user preferences, to guide LLM reasoning with a multi-stage approach to direct the final outputs. This approach is used in the LLM inference stage, reducing the need for expensive post-training, and can be adopted by UI designers for efficient prototyping and iteration in the UI design process.

\section{Formative Study} \label{sec:prelim_study}
To understand user preferences in UI interaction, we use the user-preferred UI control type out of other fundamental UI design aspects (e.g., UI control appearance and layout) as the angle. We conducted a formative study to collect user-preferred UI controls on three common UI interaction tasks. We then compared the user-preferred UI controls with LLM-generated UI controls, as well as the UI controls offered by common software tools, to surface the gap in user preference alignment.

\subsection{Study Design} \label{sec:prelim_study_design}
\hspace*{1em} \textbf{Tasks.}
We selected three common image editing tasks (\textbf{the adjustment of image lightness, saturation, and hue}) for our study since these tasks require UI interaction to handle multiple data manipulation needs, including continuous and discrete value adjustments, and need instant visual feedback for users to see the intermediate results. These requirements represent fundamental UI interaction for digital content editing, which applies to boarder tasks, such as graphic design. Additionally, image editing operations and desired results are often subjective, such that the UI controls for user interaction emphasize user preference alignment.

\textbf{User preference aspects.}
Acknowledging that there are a number of user preferences regarding UI interaction, we used \emph{predictability}, \emph{efficiency}, and \emph{explorability} as three representative user preference aspects. They came from well-established UI design principles, including McKay's ``UI is Communication''~\cite{mckay2013ui} and Lidwell et al.'s ``Universal Principles of Design''~\cite{lidwell2010universal}. Specifically, an effective UI design needs to enable effective communication with users, highlighting that an UI should be \emph{predictable} regarding its functionality, \emph{efficient} to map user operation to intended results with minimal time and effort, and \emph{explorable} to produce a wide range of potential results.

\begin{figure*}[!ht]
    \centering
    \includegraphics[trim={4.5cm 4.5cm 4.5cm 4.5cm}, clip, width=0.94\textwidth]{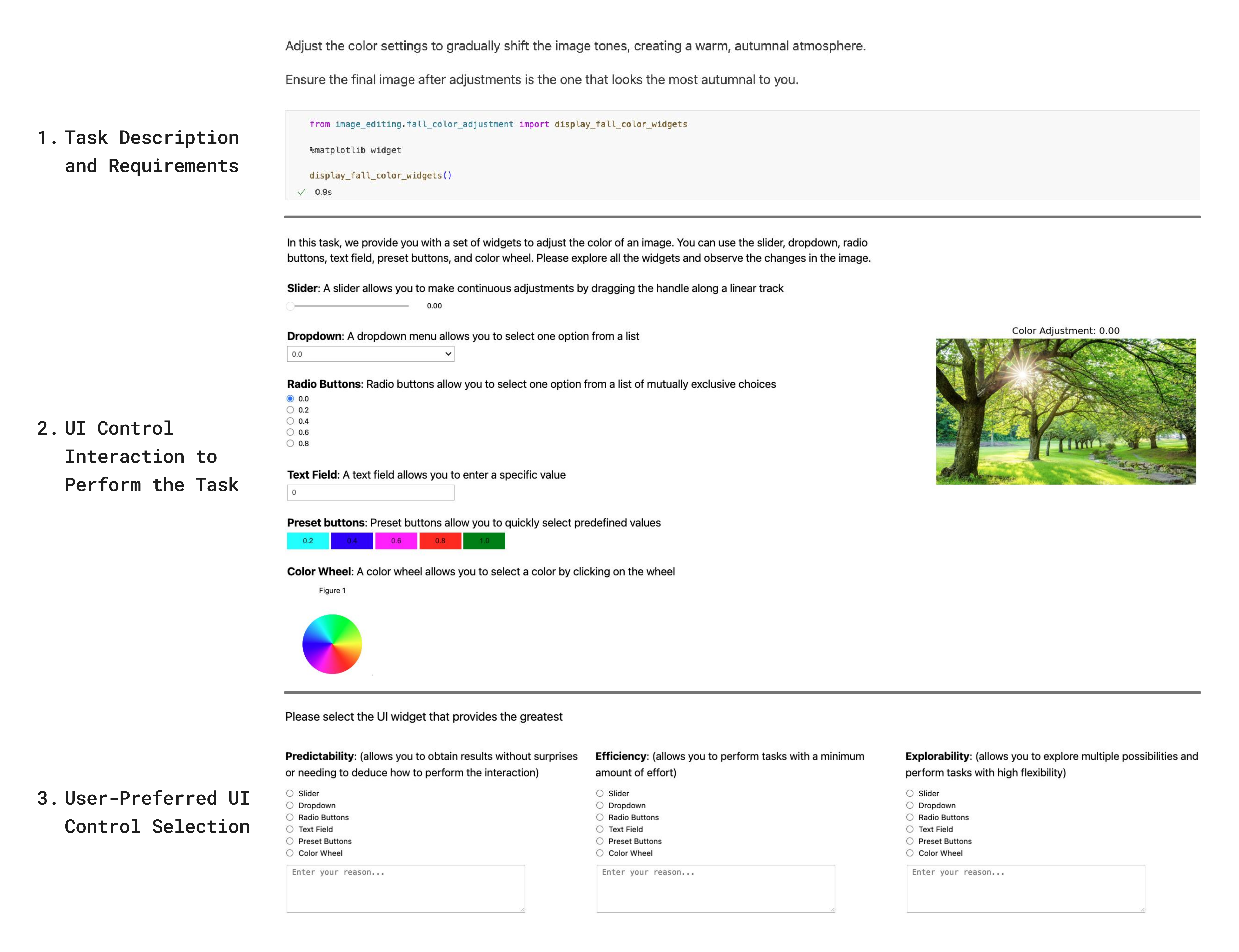}
    \caption{Example UI in our study (formative and crowdsourcing) for adjusting image hue to fall colors. It presents task description and requirements, multiple UI controls to perform the task, and a user preference selection panel for multiple user preference aspects (predictability, efficiency, explorability).}
    \Description{This figure shows an example of the crowdsourcing user interface consisting of three components. From top to bottom, the components are the task description, the UI widget interaction panel to perform the task, and the user preference selection panel. The task description specifies the goal that the user is expected to achieve. Below it, the UI widget interaction panel shows multiple widgets on the left and an image to edit on the right. At the bottom, the preference selection panel allows users to select their preferred widgets and write their reasons for the three preference aspects.}
    \label{fig:ui_image_adjust_fall_color}
\end{figure*}

\textbf{Study procedure.}
We recruited 10 participants (Age: 34.2 $\pm$ 11.2; Male: 8, Female: 2) on Prolific~\cite{prolific}. We did not require participants to have image editing or UI programming experience, as we aim to collect general users' preference data, who are the target users of our UIs. Each of them was offered a Jupyter Notebook that contained the descriptions of the three tasks, multiple UI controls to perform each task, and a user preference selection panel after each task. An example UI used in our study is shown in Figure~\ref{fig:ui_image_adjust_fall_color}. The offered UI controls support various data manipulation requirements, including continuous value adjustment (slider, text field), discrete value adjustment (drop-down menu, radio buttons, preset buttons with visual overlays), and color adjustment (color wheel). Users were instructed to interact with all the offered UI controls to perform each task and select the UI controls they preferred according to each of the three preference aspects. 

\subsection{Study Results} \label{sec:prelim_study_results}
We summarize the study results in Table~\ref{tab:prelim_pref_10}. This table compares the UI controls that the users preferred in our study, LLM (GPT-4o)-reasoned UI controls, and UI controls provided by both professional and novice-friendly image editing software. The results show that for different tasks, and for the same task but different user preference aspects, the UI controls that our participants preferred vary greatly. In addition, the user-preferred UI controls also differ from the UI controls reasoned by the LLM and those offered by widely-used image editing software (Photoshop and Adobe Express). 

\begin{table*}[!ht]
    \centering
    \caption{Comparison of user-preferred UI controls (with the number of users who preferred each type of UI control), LLM-reasoned UI controls, and UI controls provided by professional (Photoshop) and novice-oriented (Adobe Express) software.}
    \Description{This table has four columns: UI controls, adjust image lightness, adjust image saturation, and adjust image hue; and three rows: user-preferred, LLM-reasoned, and Photoshop \& Adobe Express. Each entry of the table reports the types of UI controls.}
    \scalebox{0.8} {
    \begin{tabular}{lllll}
    \toprule
        \multicolumn{2}{c}{\textbf{UI Controls}} & \textbf{Adjust Image Lightness} & \textbf{Adjust Image Saturation} & \textbf{Adjust Image Hue} \\
    \midrule
        \multirow{9}{*}{\textbf{User-Preferred}}
        & \emph{Predictability}
        & \makecell[l]{Preset buttons: 7 \\ Slider: 2 \\ Text field: 1}
        & \makecell[l]{Preset buttons: 8 \\ Slider: 2}
        & \makecell[l]{Color wheel: 7 \\ Preset buttons: 2 \\ Drop-down menu: 1} \\
        \cmidrule(l){2-5}
        & \emph{Efficiency}
        & \makecell[l]{Slider: 4 \\ Preset buttons: 3 \\ Text field: 2 \\ Radio buttons: 1}
        & \makecell[l]{Preset buttons: 6 \\ Slider: 3 \\ Radio buttons: 1}
        & \makecell[l]{Preset buttons: 5 \\ Color wheel: 3 \\ Slider: 1 \\ Radio buttons: 1} \\
        \cmidrule(l){2-5}
        & \emph{Explorability}
        & \makecell[l]{Slider: 9 \\ Text field: 1}
        & \makecell[l]{Slider: 8 \\ Text field: 2}
        & \makecell[l]{Color wheel: 8 \\ Slider: 2} \\
    \midrule
        \multirow{4}{*}{\textbf{LLM-Reasoned}} & \emph{Predictability} & Stepper input & Preset buttons & Text field \\
        \cmidrule(l){2-5}
        & \emph{Efficiency} & Text field & Stepper input & Text field \\
        \cmidrule(l){2-5}
        & \emph{Explorability} & Slider & Slider & Color wheel \\
    \midrule
        \multicolumn{2}{l}{\textbf{Photoshop \& Adobe Express}} & Slider, Text field & Slider, Text field & Slider, Text field \\
    \bottomrule
    \end{tabular}
    }
    \label{tab:prelim_pref_10}
\end{table*}

\subsection{Reflection}
\hspace*{1em} \textbf{Gap between user-preferred, LLM-reasoned, and commercial software-offered UI controls.}
From the results, we found that user-preferred UI controls show varied patterns with regard to the task and user preference aspect. These UI controls often differed from LLM-reasoned ones, even though LLMs were trained based on large-scale human data and commonsense, as well as the pre-defined UI controls in commercial software, although the software has gone through user testing and designer refinement. 

\textbf{Need for the integration of collected user preferences into UI design.}
The observed gap highlights the need for user preference-aligned UI design based on the user's task and preference. The collected user preference data motivates us to explore how it can be effectively integrated into the UI design process to assist the creation of UIs with UI controls that are aligned with target users' preferences.

\section{\crowdgenui{}} \label{sec:system}
In this section, we introduce the \crowdgenui{} method to support user preference-aligned UI design and our implementation of this method. Based on our findings in the formative study, our method is designed to fill the user preference alignment gap in UI controls and address the need to integrate user preferences into UI generation.

\subsection{Method}
\begin{figure*}[!ht]
    \centering
    \includegraphics[trim={4.5cm 4.5cm 4.5cm 4.5cm}, clip, width=\textwidth]{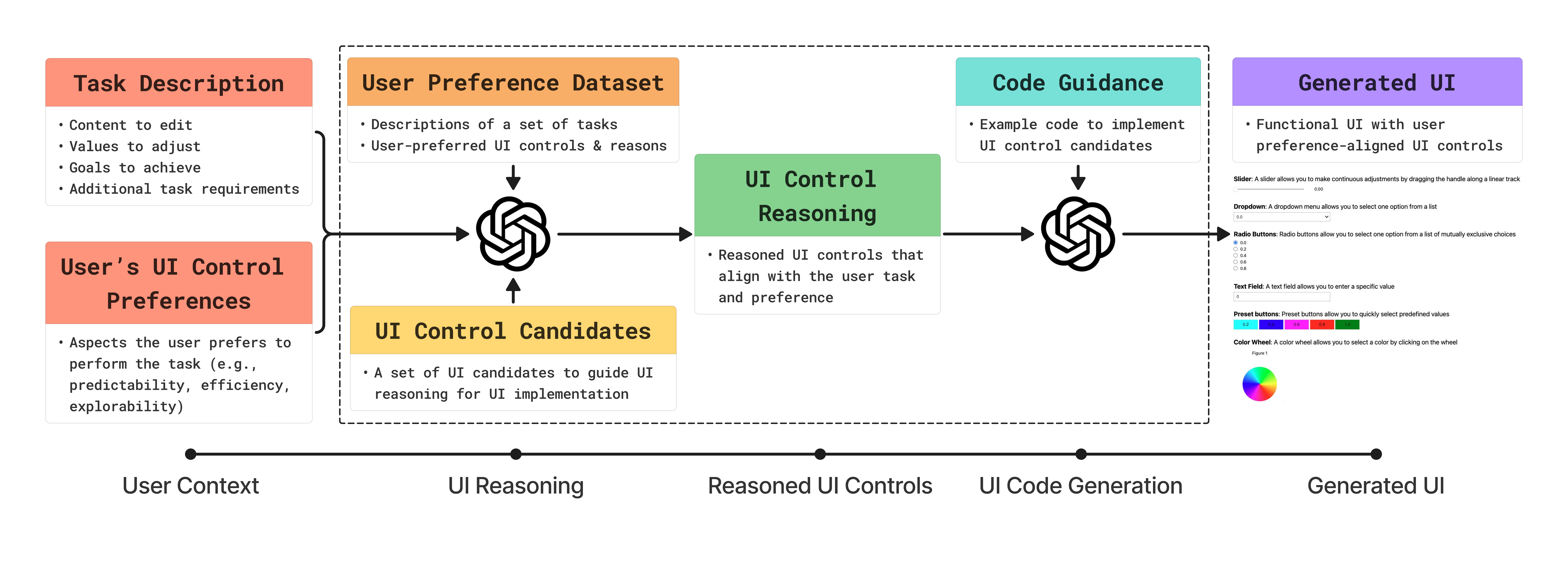}
    \caption{\crowdgenui{} method pipeline. The pipeline takes user context as input, which specifies the user's task description and UI control preferences. With this input, the LLM is additionally fed with a user preference dataset, which contains a set of tasks and user-preferred UI controls to perform these tasks regarding multiple user preference aspects, and UI control candidates, which are used to support UI control implementation. The LLM is prompted by multi-stage reasoning to obtain user preference-aligned UI controls. These reasoned UI controls are then implemented based on LLM-based code generation and presented to the user in a functional UI.}
    \Description{This figure is a flow chart showing the system implementation pipeline. Starting from the left, the task description, preference aspect, crowdsourced preference dataset, and UI widget candidates are passed as input to the LLM for multi-stage reasoning. As the reasoning output, reasoned widgets containing widget types and the corresponding rationale are returned. This output is further passed to the LLM for code generation with sample code as guidance. Finally, Python code is returned for widget display and interaction.}
    \label{fig:llm_ui_generation_pipeline}
\end{figure*}

To generate UIs that align with user preferences, \crowdgenui{} takes user context as input, which consists of the user's task and preference, and directs LLMs to reason and implement user preference-aligned UI controls guided by a user preference dataset. Figure~\ref{fig:llm_ui_generation_pipeline} illustrates the \crowdgenui{} method pipeline.

\subsubsection{User Context as Input}
Our method takes user context as input, which consists of the user's task description and UI control preferences. 

\textbf{Task description.}
The task description in natural language specifies the content users want to edit (e.g., an image), values to adjust (e.g., image color), goals to achieve (e.g., changing the image to fall colors), and any additional user requirements to complete the task. 

\textbf{UI control preferences.} \label{sec:user_preference_aspects}
After specifying what to achieve in the task description, users can add what aspects they prefer when performing the task as UI control preferences. This can include multiple preference aspects, such as high-efficiency content editing or wide-explorability value adjustments. Users describe these preferences in natural language. 

\subsubsection{UI Reasoning according to User Context}
The user context (user task and preference) is passed to the LLM for UI generation reasoning. In addition, a user preference dataset and UI control candidates are fed to the LLM to guide the multi-stage reasoning process to reason UI controls that align with user preferences. 
Specifically, the LLM first needs to analyze the user's task and identify the task requirements. Based on this analysis, the LLM is guided to analyze the similarity between the user task and the tasks in the user preference dataset by comparing task descriptions. The similarity is compared based on the fundamental value adjustment requirement involved in the task, e.g., tasks that require continuous value adjustment are considered similar, while tasks that need continuous or discrete value adjustment are not. 
Once similar tasks are identified, the LLM is prompted to look for user-preferred UI controls in each relevant task from the dataset according to the corresponding preference aspect. Since we found that the LLM can accidentally reason UI controls that are out of the scope of the user preference dataset, to make sure the reasoned UI controls can be successfully implemented, the LLM needs to conduct a sanity check on the reasoned UI controls and only keep the ones that belong to the UI control candidates. 
Finally, the UI reasoning concludes by returning a JSON output that contains the user task, user-specified preference aspect, user preference-aligned UI controls, and LLM reasoning that describes why the reasoned UI controls align with user preferences for the user's task. 
This UI reasoning process is outlined in Algorithm~\ref{algo:cot_reasoning} and the full prompts can be found in Appendix~\ref{appendix:prompt_widget_reasoning_withlib}.

\begin{algorithm*}
\caption{UI Reasoning based on User Task and Preference}
\label{algo:cot_reasoning}

\begin{algorithmic}[0]
\STATE \textbf{Input}: \textcolor{black}{user-task-description} ($\mathcal{T}$), \textcolor{black}{user-preference-aspect} ($\mathcal{A}$), \textcolor{black}{{user-preference-dataset}} ($\mathcal{D}$), \textcolor{black}{ui-control-candidates} ($\mathcal{C}$)
\STATE \textbf{Output}: \textcolor{black}{ui-control-reasoning}
\end{algorithmic}

\begin{algorithmic}[1]
\FOR{\textcolor{black}{each} \textcolor{black}{task} \textcolor{black}{in} $\mathcal{D}$}
    \STATE \textcolor{black}{relevant-tasks} $\gets$ \textit{similarity} ($\mathcal{T}$, \textcolor{black}{{task}}) \COMMENT{Identify tasks in the user preference dataset that are similar to the user task}
    
    \FOR{\textcolor{black}{each} \textcolor{black}{aspect} \textcolor{black}{in} $\mathcal{A}$}
        \FOR{\textcolor{black}{each} \textcolor{black}{relevant-aspect} \textcolor{black}{in} \textcolor{black}{relevant-tasks}}
            \STATE \textcolor{black}{relevant-task-aspect} $\gets$ \textit{match} (\textcolor{black}{relevant-aspect}, \textcolor{black}{aspect}) \COMMENT{Identify user preferences that match}
            \STATE \textcolor{black}{ui-controls} $\gets$ \textit{match} (\textcolor{black}{relevant-task-aspect}, \textcolor{black}{{aspect}}) \COMMENT{Find UI controls that align with user preferences}
            \RETURN \textcolor{black}{ui-controls} $\in \mathcal{C}$ \COMMENT{Return user preference-aligned UI controls that belong to the UI control candidates for successful implementation}
        \ENDFOR
    \ENDFOR
\ENDFOR

\RETURN \textcolor{black}{ui-control-reasoning} $\gets$ JSON (\textcolor{black}{ui-controls}) \COMMENT{Organize results in JSON format, which contain user task, user-specified preferences, user preference-aligned UI controls, and LLM reasoning}
\end{algorithmic}
\end{algorithm*}

\subsubsection{UI Generation with LLM-Generated Code}
Once the reasoned UI controls are obtained, the LLM is further guided to generate the code to implement these controls. During this process, \crowdgenui{} provides an example code, which specifies the coding language and covers the implementation of all the UI control candidates as the guidance for code generation. Based on our experiments, the code guidance can greatly reduce syntax errors in LLM-generated code to produce functional UIs. The detailed prompts and code guidance for UI control implementation can be found in Appendix~\ref{appendix:prompt_widget_coding}. 

\subsection{Implementation}
Here, we present our implementation details of the \crowdgenui{} method. 

\subsubsection{UI Implementation}
The LLM-based UI reasoning and UI code generation have been tested with GPT-4o~\cite{gpt4o}. 
We implement the LLM-generated UIs in Jupyter Notebook as our testbed since it provides a wide range of UI controls from the Jupyter Widgets Library~\cite{ipywidgets} and allows easy user interaction. Although this implementation does not cover other types of UIs (e.g., web, desktop, and mobile), our UI generation method is extendable to accommodate these UIs by generating UI implementation code in the corresponding programming languages. In addition, our work currently focuses on the reasoning and implementation for UI controls, which are the fundamental components in UI design. As introduced in Section~\ref{sec:relatedwork}, the layout and appearance of UI controls also play important roles in UI design, and the idea to reason and implement UI controls follows a similar logic to the rest aspects in UI design. 

\subsubsection{User Preference Data Collection.} \label{sec:crowdsourcing}
Motivated by using crowdsourcing for user feedback collection to assist various design areas~\cite{xu2014voyant, hossain2015crowdsourcing, xu2015classroom, luther2015structuring, yu2016encouraging, foong2017novice, lee2018exploring, oppenlaender2020crowdui}, we use crowdsourcing as our method to collect the user preference dataset. This method aligns with our goal to collect user preference data from general users, who are the target users of many UI designs. The data collection method can be adapted to support different user groups, such as UI designers, professionals with specialized UI needs, or individuals requiring personalized UI design. 

In the following sections, we introduce the details of our crowdsourcing setup and analysis of the crowdsourced data.

\paragraph{Setup}
Here, we introduce the crowdsourcing setup, including tasks, participant recruitment, UI control candidates, and crowdsourcing procedure. 

\begin{table*}[!ht]
    \centering
    \caption{Tasks used in the crowdsourcing to collect our user preference dataset.}
    \Description{This table consists of two columns: task name and task description. In each row, the number, name, and description of each crowdsourcing task are detailed.}
    \scalebox{0.8} {
    \begin{tabular}{cp{0.35\textwidth}p{0.65\textwidth}}
    \toprule
         & \multicolumn{0}{c}{\textbf{Task Name}} & \multicolumn{0}{c}{\textbf{Task Description}}\\ \midrule
        1 & \texttt{image\_adjust\_lightness} & Experiment with lightness settings to see how different levels affect the image.\\ \hline
        2 & \texttt{image\_adjust\_saturation} & Boost the saturation to make the colors pop.\\ \hline
        3 & \texttt{image\_adjust\_hue} & Experiment with different hues to find a color tone that complements the overall mood of the image.\\ \hline
        4 & \texttt{image\_adjust\_fall\_color} & Adjust the color settings to gradually shift the image tones, creating a warm, autumnal atmosphere. Ensure the final image is the one that looks the most autumnal to you.\\ \hline
        5 & \texttt{image\_color\_match} & Change the rocket's color to match the provided reference color. Ensure the rocket's color in the final image best matches the reference color.\\ \hline
        6 & \texttt{image\_adjust\_color\_balance} & Experiment with different color balance settings to see how altering the red, green, and blue levels affects the overall color harmony of the image.\\ \hline
        7 & \texttt{image\_place\_watermark} & Experiment with different watermark positions to find the perfect balance between visibility and subtlety.\\ \hline
        8 & \texttt{image\_place\_vignette} & Darken the background using a vignette effect except for the human face by properly positioning the circle around the face.\\
    \bottomrule
    \end{tabular}
    }
    \label{tab:crowdsourcing_tasks}
\end{table*}

\textbf{Tasks.} \label{sec:crowdsourcing_tasks}
Similar to the formative study, we used image editing tasks for the crowdsourcing. Specifically, we used the three tasks from the formative study and additionally included five tasks (i.e., eight tasks in total). Detailed task descriptions are available in Table~\ref{tab:crowdsourcing_tasks}. These tasks require \textbf{continuous value adjustment} (Tasks 1, 2, 3, 6), \textbf{discrete value selection} (Tasks 1, 2, 3, 7), and \textbf{color/position adjustment} (Tasks 4, 5, 7, 8). We used these tasks to encourage either outcome exploration (Tasks 1, 2, 3, 6, 7) or operation precision (Tasks 4, 5, 8). Among them, Tasks 4 and 5 were used as control tasks for attention check (the final edited images should meet the task requirements) to help us identify untrustworthy responses. The order of tasks was randomized across users. 

\textbf{Participants.}
Following the same participant recruitment setup as the formative study, we recruited participants on Prolific. They were general users on image editing and UI programming. According to prior research on user testing for UX design~\cite{tomczak2023over, usertesting2025sample, userinterviews2025preference}, 20–30 participants are generally sufficient to evaluate simple to intermediate tasks, and 50 participants are recommended for complex tasks. Following these guidelines, we invited 50 participants in total (Age: 29.86 $\pm$ 7.83; Male: 37, Female: 12, Non-binary: 1), and among them, 10 were from the formative study. These 10 users worked on the five additional tasks (Tasks 4-8) and their responses to Tasks 1-3 from the formative study were included into our collected data.

\textbf{UI control candidates.} \label{sec:widget_selection_implementation}
In addition to the UI controls used in the formative study (Section~\ref{sec:prelim_study_design}), we added the color picker for color adjustment tasks (Tasks 5, 6) and direct manipulation by clicking on the image for position adjustment tasks (Tasks 7, 8). The full set of UI control candidates contains: \emph{slider}, \emph{text field}, \emph{drop-down menu}, \emph{radio buttons}, \emph{preset buttons} with preview overlays, \emph{color wheel}, \emph{color picker}, and direct manipulation (e.g., \emph{clicking}). 
Although these candidates are selected to support various types of UI interaction, a defined set might limit other UI control possibilities. Based on our experiments, removing these UI control candidates would sometimes yield LLM-reasoned UI controls being unimplementable as these reasoned UI controls can go beyond the Jupyter Widgets Library. From the implementation perspective, it is challenging to go infinite on the UI control candidates, but this candidate pool can be extended to allow broader UI interaction needs. 

\textbf{Crowdsourcing procedure.}
The crowdsourcing followed the same procedure (task description, UI control interaction, user preference selection) as the formative study. We distributed the crowdsourcing tasks in three versions. Each version included all eight tasks and a subset of the three preference aspects (predictability, efficiency, explorability), covering one, two, or all three aspects. This approach ensured that each participant provided their preferences for 15-18 preference selections out of the total 3 aspects $\times$ 8 tasks = 24 selections to prevent fatigue. The study yielded a user preference dataset with 30 unique user responses for each preference aspect.

\paragraph{User Preference Data} \label{sec:crowdsourcing_result_analysis}
Next, we summarize the crowdsourced data and analyze the collected user preferences. 

\begin{figure*}[!ht]
    \centering
    \includegraphics[width=0.9\textwidth]{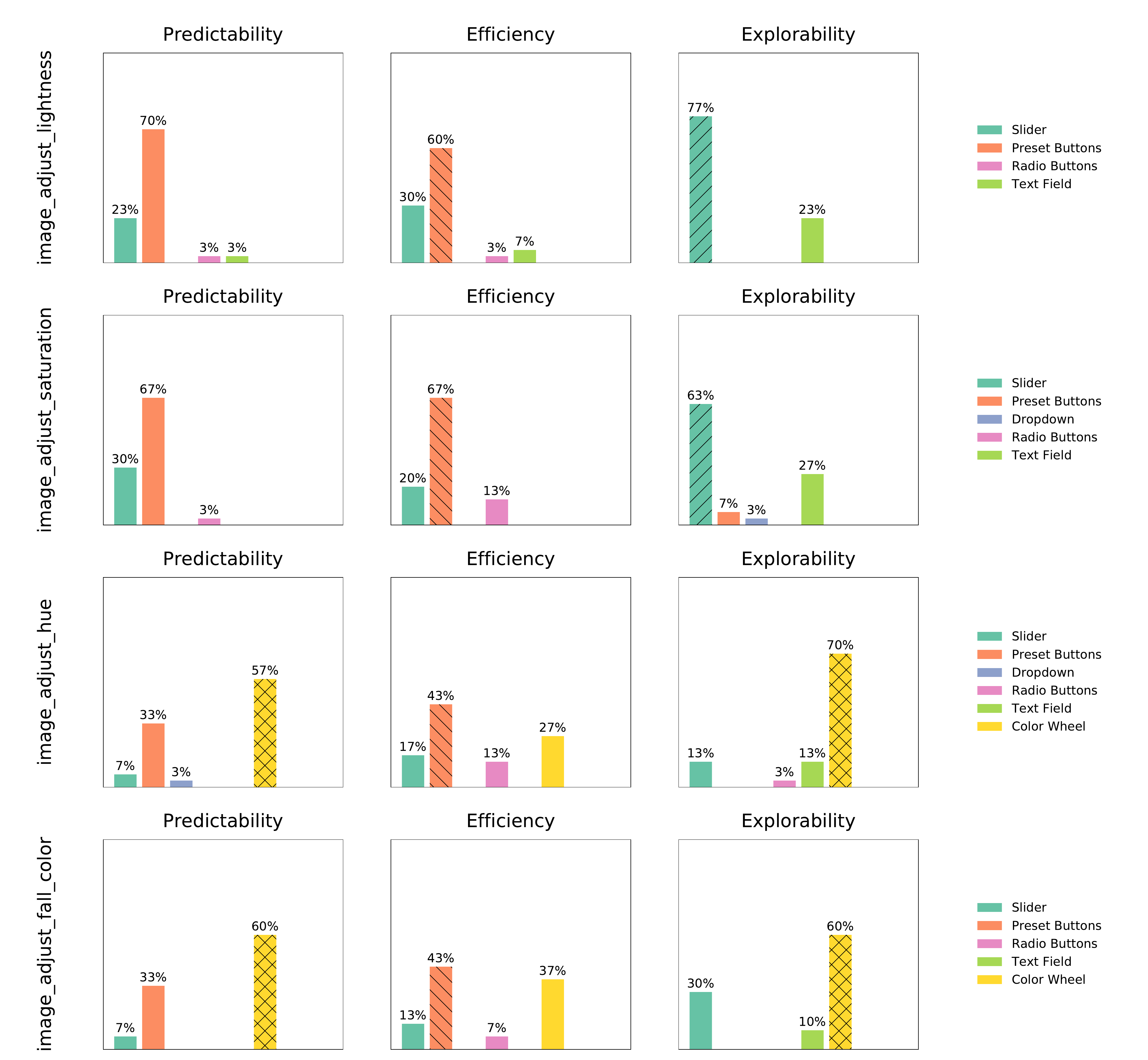}
    \caption{Crowdsourcing results 1. User preference distribution of tasks 1-4 (task set 1).}
    \Description{This figure consists of four rows, showing the crowdsourcing results of tasks 1 to 4 in each row. Each row has three columns, depicting the user preference distribution of the three preference aspects in each column as bar charts.}
    \label{fig:crowdsourcing_preferences_1to4}
\end{figure*}

\begin{figure*}[!ht]
    \centering
    \includegraphics[width=0.9\textwidth]{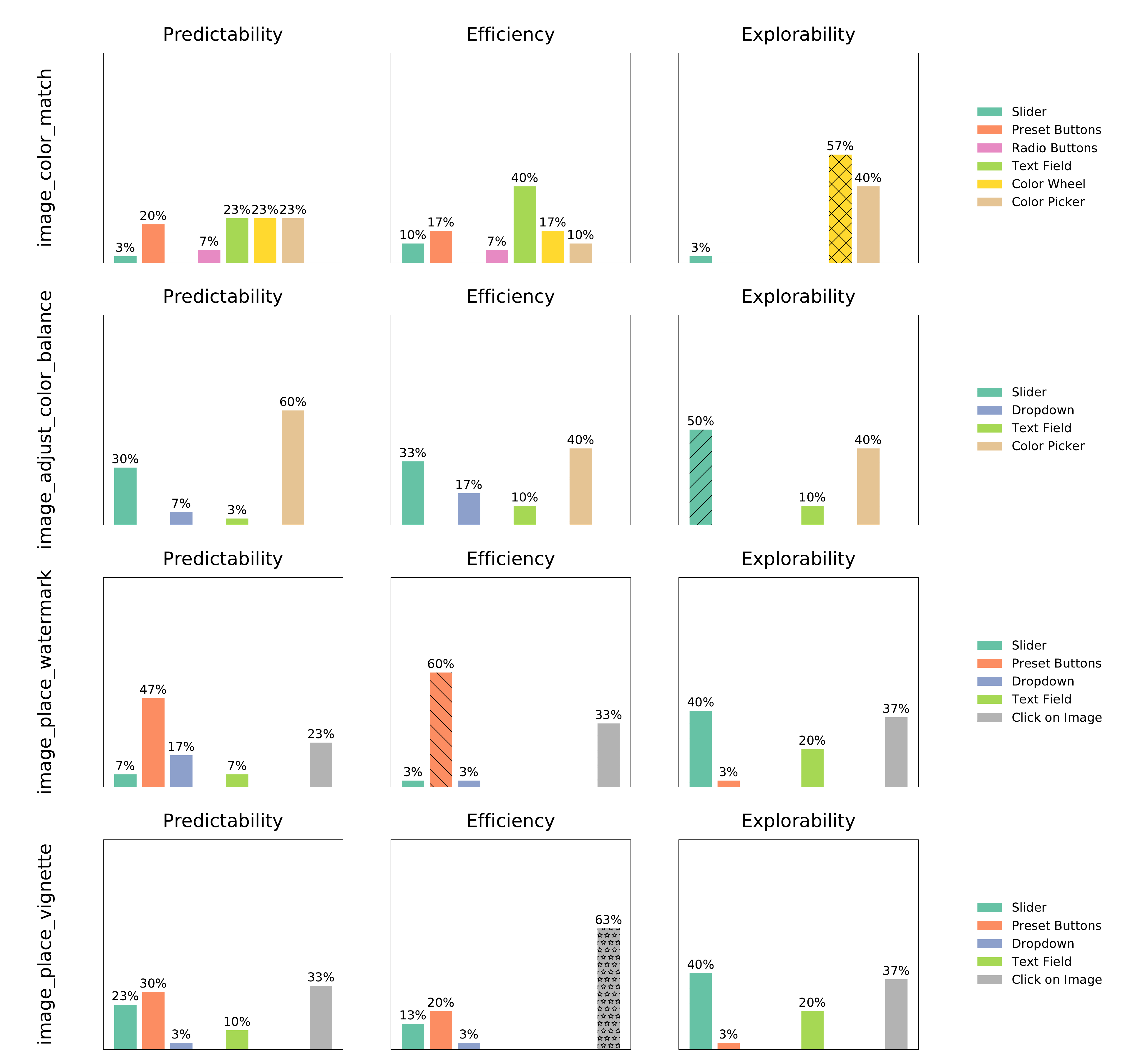}
    \caption{Crowdsourcing results 2. User preference distribution of tasks 5-8 (task set 2).}
    \Description{This figure consists of four rows, showing the crowdsourcing results of tasks 5 to 8 in each row. Each row has three columns, depicting the user preference distribution of the three preference aspects in each column as bar charts.}
    \label{fig:crowdsourcing_preferences_5to8}
\end{figure*}

\textbf{Statistics overview.}
In total, we obtained 30 responses $\times$ 3 aspects $\times$ 8 tasks = 720 pieces of user preferences. Figures~\ref{fig:crowdsourcing_preferences_1to4} and~\ref{fig:crowdsourcing_preferences_5to8} show the statistics overview of collected user preferences.

\textbf{Analysis of user preferences.}
From the results, we observe variations in user preferences for UI controls on predictability, efficiency, and explorability with regard to different task requirements:

\begin{itemize}[left=0pt]
    \item \textbf{Continuous value adjustment.} Users preferred preset buttons with preview overlays for predictability and efficiency. For explorability, sliders were favored for adjusting lightness and saturation, while color wheels were preferred to explore hue adjustment.

    \item \textbf{Discrete value selection.} Preset buttons were generally favored over radio buttons and drop-down menus in all three preference aspects, due to the provided visual overlays.

    \item \textbf{Color/position adjustment.} (1) In color adjustment tasks, many users preferred color wheels and color pickers for predictability and explorability, while preset buttons were preferred for efficiency. Additionally, for precise color adjustment (e.g., matching an object's color to a given color), no clear consensus was obtained regarding predictability, where opinions were evenly split among text fields, color wheels, color pickers, and preset buttons. For efficiency, most users preferred text fields, and for explorability, color wheels and color pickers were the most favored. (2) In position adjustment tasks, preset buttons and directly clicking on the image were preferred for predictability and efficiency, while for explorability, users favored both clicking on the image and using sliders.
\end{itemize}

\begin{figure}[!ht]
    \centering
    \includegraphics[width=\linewidth, trim={4.5cm 4.5cm 4.5cm 4.5cm}, clip]{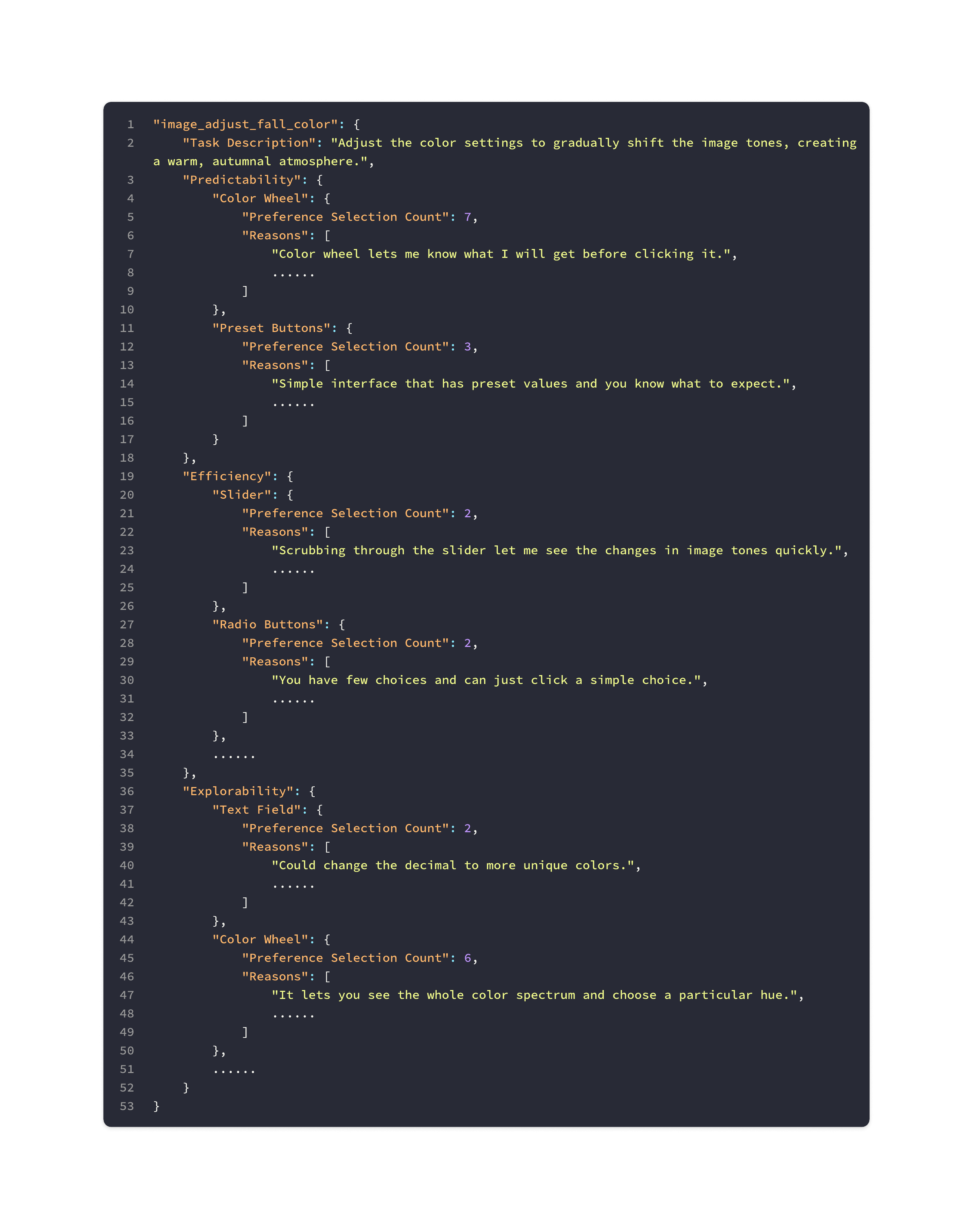}
    \caption{Example of crowdsourced user preference data for task \texttt{image\_adjust\_fall\_color}. The data is organized in the JSON format in our user preference dataset.}
    \Description{This image shows the user preference dataset in the JSON format, specifically for task adjusting an image to fall colors.}
    \label{fig:user_dataset_example}
\end{figure}

\textbf{Data organization for the user preference dataset.}
We organize the crowdsourced user preference data in the JSON format. The content is structured as the layers of \emph{task}, \emph{user preference aspects}, and \emph{user-preferred UI controls}. Specifically, the \emph{task} contains the task description and requirements, and the \emph{user preference aspects} include the three preference aspects. In each preference aspect, it summarizes the \emph{user-preferred UI controls}, which consist of the count of user-preferred UI controls and the corresponding user-provided reasons. In Figure~\ref{fig:user_dataset_example}, we show an example of the organized JSON data for task \texttt{image\_adjust\_fall\_color}.

\section{User Study} \label{sec:evaluation}
To evaluate the UIs generated using our \crowdgenui{} method, we conducted a user study. In this section, we introduce our research questions, study design and results.

\subsection{Research Questions} \label{sec:research_questions}
We explore the following research questions (RQs) to evaluate the effectiveness of integrating the user preference dataset into UI generation:

\begin{itemize}[left=0pt]
    \item \textbf{RQ1: Generalizability of the \crowdgenui{} method to generate UIs for new tasks.} In the user study, we use tasks that differ from but relate to the crowdsourcing tasks to explore the method's generalizability.

    \item \textbf{RQ2: Impact of the user preference dataset on generated UIs regarding user preference alignment.} We compare user feedback on the generated UIs with and without the user preference dataset to understand the dataset's impact on user preference alignment. 

    \item \textbf{RQ3: Effect of user preference dataset sizes on the generated UIs.} Since collecting data takes time and effort, we explore the tradeoff between dataset size and user preference alignment to derive insights on the effective size of such a dataset for LLM-generated UI alignment with user preferences. 
\end{itemize}

Next, we present the user study design and analyze the results from the lens of these research questions.

\subsection{Study Design}
\subsubsection{Tasks}
\label{sec:user_eval_tasks_aspects}
The user study tasks differ from the crowdsourcing tasks but still fall within the same three categories as outlined below. The task descriptions can be found in Table~\ref{tab:user_eval_tasks}. 

\begin{itemize}[left=0pt]
    \item \textbf{Continuous value adjustment.} Adjust image exposure, tint, and temperature. 
    \item \textbf{Discrete value selection.} Choose predefined values for image exposure, tint, and temperature. Align text and position a logo.
    \item \textbf{Color/position adjustment.} Change image color to spring colors, align text with a margin, and position a logo.
\end{itemize}

\begin{table*}[!ht]
    \centering
    \caption{Tasks in the user study, which differ from the crowdsourcing tasks.}
    \Description{This table consists of two columns: task name and task description. In each row, the number, name, and description of each user evaluation task are detailed.}
    \scalebox{0.8} {
    \begin{tabular}{cp{0.35\textwidth}p{0.65\textwidth}}
    \toprule
         & \multicolumn{0}{c}{\textbf{Task Name}} & \multicolumn{0}{c}{\textbf{Task Description}}\\ \midrule
        1 & \texttt{image\_adjust\_exposure} & Adjust the image exposure by both decreasing and increasing it. Find the exposure level that appears the most natural and visually pleasing to you.\\ \hline
        2 & \texttt{image\_adjust\_tint} & Adjust the image tint to shift its color balance, creating both subtle and dramatic effects by adding a yellow or magenta hue.\\ \hline
        3 & \texttt{image\_adjust\_temperature} & Adjust the image temperature to warm and cool tones. Experiment with different settings to achieve the most desired effect for you.\\ \hline
        4 & \texttt{image\_change\_to\_spring} & Transform the image to reflect vibrant spring colors, adding fresh, lively hues.\\ \hline
        5 & \texttt{design\_align\_text} & Align the text ``Poster Title'' perfectly along one of the margins. Experiment with various positions to achieve a balanced and visually pleasing design.\\ \hline
        6 & \texttt{design\_position\_logo} & Experiment with various placements for the logo to determine the most visually appealing position that enhances visibility and complements the overall design of the image.\\
    \bottomrule
    \end{tabular}
    }
    \label{tab:user_eval_tasks}
\end{table*}

These tasks are designed to probe the \crowdgenui{} method's generalizability beyond the crowdsourcing scope to answer \textbf{RQ1}. Although details are different, the tasks fundamentally require continuous and discrete value manipulation and visual feedback. In addition, these tasks balance objective precision (e.g., margin alignment) and open-ended exploration (e.g., change to spring colors). These UI interaction requirements apply to the image editing domain, as well as broader digital content editing tasks. 

\subsubsection{Participants}
We conducted the user study on Prolific and employed the same recruitment criteria as the formative and crowdsourcing studies to evaluate the generated UI alignment with general users' preferences. None of the recruited users participated in our formative or crowdsourcing studies. In total, we recruited 78 participants (Age: 29.81 $\pm$ 9.09; Male: 56, Female: 19, Non-binary: 3) and compensated each of them 10 USD for the 30-minute study. 

\subsubsection{Procedure}
After participants responded to our study call and offered us their consent, they accessed the study via a Jupyter Notebook. Before starting the tasks, they were given a briefing that outlined the study requirements and data collection. Full details of the user study instructions can be found in Appendix~\ref{appendix:user_eval_briefing_instructions}. Following that, each of them worked on three out of the six tasks, compared UI controls for each task, and selected their preferences. We introduce how the tasks were distributed in Section~\ref{sec:pairwise_comparison_setup} and Table~\ref{tab:pairwise_comparisons}. 

\begin{figure*}[!ht]
    \centering
    \includegraphics[trim={4.5cm 4.5cm 4.5cm 4.5cm}, clip, width=0.94\textwidth]{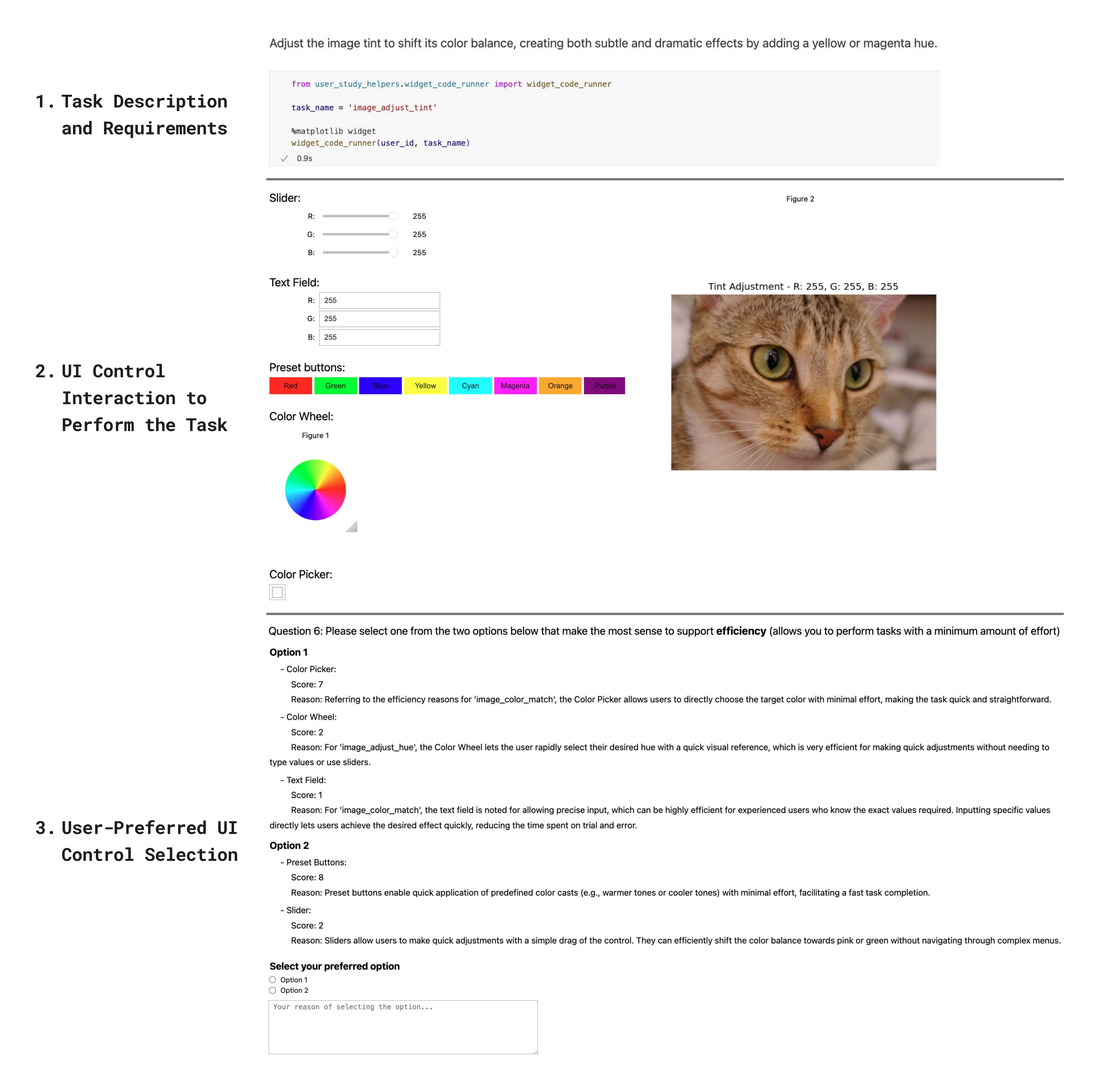}
    \caption{Example user study UI for task \texttt{image\_adjust\_tint}. The preference selection panel contains six questions for each task, and we show one of them in this figure due to the space limit.}
    \Description{This figure shows an example of the user evaluation UI consisting of three components. From top to bottom, the components are the task description, the UI widget interaction panel to perform the task, and the user preference selection panel. The task description specifies the goal that the user is expected to achieve. Below it, the UI widget interaction panel shows multiple widgets on the left and an image to edit on the right. At the bottom, the preference selection panel allows users to select their preferred widget option and write their reasons for a certain preference aspect.}
    \label{fig:ui_image_adjust_tint}
\end{figure*}

Figure~\ref{fig:ui_image_adjust_tint} shows an example task (adjusting image tint). Participants interacted with the UI, which shows the task description and requirements, UI controls to perform the task, and a preference selection panel. They needed to use all the UI controls to perform their tasks and pick their preferred UI controls with regard to the specified preference aspect. These UI controls were generated by the LLM guided by no user preference dataset or by user preference datasets of different sizes. How these UI controls were generated was hidden from the participants to prevent bias. In the next section, we introduce the details of these UI control comparisons.

\subsubsection{UI Control Comparison} \label{sec:userstudy_ui_control_comp}
To evaluate the impact of the existence (\textbf{RQ2}) and size (\textbf{RQ3}) of user preference dataset on the generated UIs, we employ four conditions for comparison: \texttt{withoutpref} (with no user preference, i.e., LLM alone), \texttt{withpref10} (with 10 pieces of user preferences), \texttt{withpref25} (with 25 pieces of user preferences), and \texttt{withpref30} (with 30 pieces of user preferences). The full user preference dataset contains 30 pieces of user preferences, and we randomly selected 10 and 25 preference subsets to obtain \texttt{withpref10} and \texttt{withpref25}.

\begin{figure*}[!ht]
    \centering
    \includegraphics[width=0.95\textwidth]{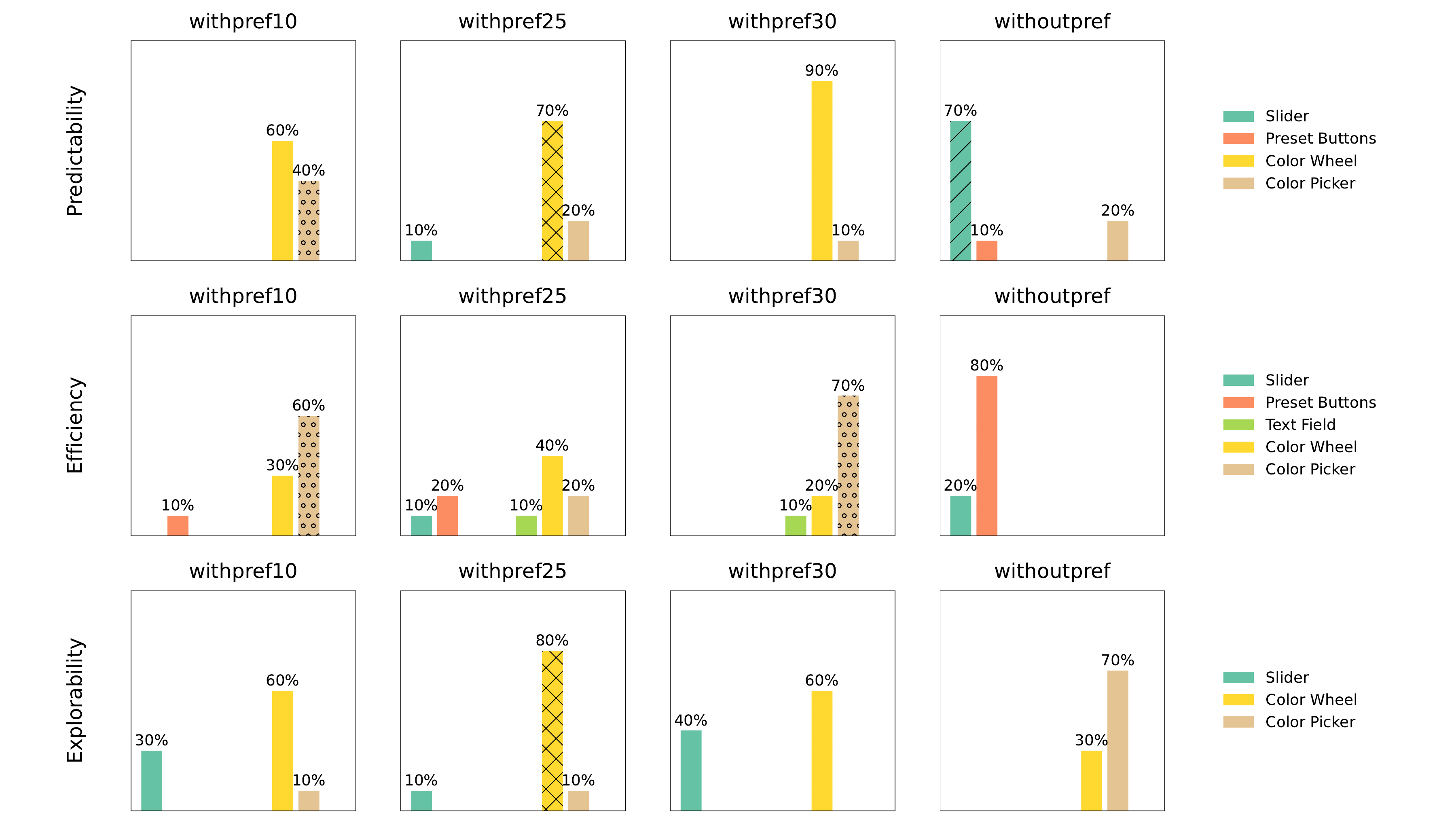}
    \caption{LLM-reasoned UI controls for task \texttt{image\_adjust\_tint} with three sizes of the crowdsourced user preference datasets (\texttt{withpref10}, \texttt{withpref25}, \texttt{withpref30}) and with no user preference dataset (\texttt{withoutpref}).}
    \Description{This figure consists of three rows, showing the LLM-reasoned widgets for the three preference aspects in each row. Each row has four columns, depicting the reasoned widget frequency as percentages using the four dataset sizes---withpref10, withpref25, withpref30, withoutpref---in each column as bar charts.}
    \label{fig:reasoning_image_adjust_tint_main}
\end{figure*}

\textbf{Impact of user preference dataset conditions on LLM generated UI controls.}
We present an example in Figure~\ref{fig:reasoning_image_adjust_tint_main} to show LLM-generated UI controls for task \texttt{image\_adjust\-\_tint} under the four conditions, and Appendix~\ref{appendix:widgets_of_user_evaluation_tasks} presents the LLM-generated UI controls for all six tasks. 
The percentages show the frequency of each type of UI control reasoned by the LLM from 10 runs. For instance, in the bar chart of \texttt{withoutpref} Efficiency (the rightmost chart in the second row of Figure~\ref{fig:reasoning_image_adjust_tint_main}), the preset buttons are reasoned 8 times (80\%), while the slider is reasoned 2 times (20\%). 
From these results, we observe that the existence of the user preference dataset had a clear impact on LLM-reasoned UI controls (\textbf{RQ2}) since disagreements between \texttt{withoutpref} and the rest three conditions occur in many tasks. In addition, disagreements also exist for different user preference dataset sizes (\textbf{RQ3}), e.g., in Figure~\ref{fig:reasoning_image_adjust_tint_main} Efficiency (the second row), \texttt{withpref10} and \texttt{withpref30} identified ``Color Picker'' as the most aligned with user preference, while \texttt{withpref25} suggested ``Color Wheel'' instead. 

To take into account the inconsistent nature of LLM outputs~\cite{barke2023grounded}, we use the frequency of LLM-reasoned UI controls to indicate the \emph{weight} of the UI controls' user preference alignment. For example, the frequencies of 8 for preset buttons and 2 for the slider are used as weights 8 and 2 for them, as seen in Figure~\ref{fig:ui_image_adjust_tint} Option 2, indicating that preset buttons are reasoned as more aligned with user preferences than the slider. This setup reflects the user preference dataset's impact on both the type and recommendation weight of UI controls and invites user assessment of such impact regarding user preference alignment. 

\textbf{Setup to compare different user preference dataset conditions' impact.} \label{sec:pairwise_comparison_setup}
We perform pairwise comparisons for LLM generated UI controls when \texttt{withpref10}, \texttt{withpref25}, \texttt{withlib\-30}, and \texttt{with\-outpref} were used. For each task, users need to select their preferences based on a comprehensive consideration of the UI controls, their weights (reasoned frequencies), and the LLM's reasoning (explains why each UI control was reasoned as aligned with the task and user preference aspect). From these comparisons, each preference aspect in each task results in 6 pairwise comparisons, as outlined in Table~\ref{tab:pairwise_comparisons}.

\begin{table*}[!ht]
    \centering
    \caption{Pairwise comparisons of UI controls generated based on the four user preference dataset conditions (\texttt{withpref10}, \texttt{withpref25}, \texttt{withlib\-30}, and \texttt{withoutpref}). Pair order randomized in the user study.}
    \Description{This table details the six possible pairwise comparisons of dataset sizes.}
    \scalebox{0.8} {
        \begin{tabular}{cl| cl |cl}
        \toprule
        \multicolumn{6}{c}{\textbf{User Preference Dataset Size Comparison}} \\
        \midrule
        1 & \texttt{withpref10} \textit{vs.} \texttt{withpref25}
        &
        2 & \texttt{withpref10} \textit{vs.} \texttt{withpref30}
        &
        3 & \texttt{withpref10} \textit{vs.} \texttt{withoutpref}
        \\
        \midrule
        4 & \texttt{withpref25} \textit{vs.} \texttt{withpref30}
        &
        5 & \texttt{withpref25} \textit{vs.} \texttt{withoutpref}
        &
        6 & \texttt{withpref30} \textit{vs.} \texttt{withoutpref}
        \\
        \bottomrule
        \end{tabular}
    }
\label{tab:pairwise_comparisons}
\end{table*}

In total, this gives us 6 pairs $\times$ 3 aspects $\times$ 6 tasks = 108 comparisons. To prevent user fatigue, each participant was assigned 18 pairs, which consisted of all six pairs for each preference aspect across three tasks. Specifically, the tasks were divided into two task sets, each containing three tasks. Each participant provided preferences for all three aspects of their assigned task set. The tasks were presented in all six permutations to account for order effects, and the sequence of aspects was counterbalanced using the Latin Square method~\cite{colbourn2010crc}. See Table~\ref{tab:user_eval_setup} for this setup. 

\begin{table*}[!ht]
    \centering
    \caption{User study task setup. The six user evaluation tasks are divided into two task sets. Within each set, all permutations of task and aspect orders are employed to control for order effects, and the sequence of pairwise comparisons is randomized.}
     \Description{This table consists of two components: the evaluation setup for task sets 1 and 2. In each component, the sub-table contains three columns: the total order number of tasks, task names, and preference aspects.}
    \scalebox{0.8} {
    \begin{tabular}{cccccc}
    \toprule
        \multicolumn{3}{c}{\textbf{Task Set 1}} & \multicolumn{3}{c}{\textbf{Task Set 2}} \\
        \cmidrule(r){1-3} \cmidrule(l){4-6}
        \textit{Order} & \textit{Task Name} & \textit{Aspect} & \textit{Order} & \textit{Task Name} & \textit{Aspect} \\ 
        \cmidrule(r){1-3} \cmidrule(l){4-6}
        \multirow{4}{*}{6 $\times$} & \texttt{image\_adjust\_exposure} & Predictability & \multirow{4}{*}{6 $\times$} & \texttt{image\_adjust\_tint} & Predictability\\
        \cmidrule(r){2-3} \cmidrule{5-6}
         & \texttt{image\_adjust\_temperature} & Efficiency &  & \texttt{image\_change\_to\_spring} & Efficiency\\
        \cmidrule(r){2-3} \cmidrule{5-6}
         & \texttt{design\_align\_text} & Explorability &  & \texttt{design\_position\_logo} & Explorability\\ 
        \cmidrule(r){1-3} \cmidrule(l){4-6}
        \multirow{4}{*}{6 $\times$} & \texttt{image\_adjust\_exposure} & Efficiency & \multirow{4}{*}{6 $\times$} & \texttt{image\_adjust\_tint} & Efficiency\\
        \cmidrule(r){2-3} \cmidrule{5-6}
         & \texttt{image\_adjust\_temperature} & Explorability &  & \texttt{image\_change\_to\_spring} & Explorability\\
        \cmidrule(r){2-3} \cmidrule{5-6}
         & \texttt{design\_align\_text} & Predictability &  & \texttt{design\_position\_logo} & Predictability\\ 
        \cmidrule(r){1-3} \cmidrule(l){4-6}
        \multirow{4}{*}{6 $\times$} & \texttt{image\_adjust\_exposure} & Explorability & \multirow{4}{*}{6 $\times$} & \texttt{image\_adjust\_tint} & Explorability\\
        \cmidrule(r){2-3} \cmidrule{5-6}
         & \texttt{image\_adjust\_temperature} & Predictability &  & \texttt{image\_change\_to\_spring} & Predictability\\
        \cmidrule(r){2-3} \cmidrule{5-6}
         & \texttt{design\_align\_text} & Efficiency &  & \texttt{design\_position\_logo} & Efficiency\\
    \bottomrule
    \end{tabular}
    }
    \label{tab:user_eval_setup}
\end{table*}

\textbf{Data analysis.}
We employ the Chi-squared ($\chi^2$) test~\cite{pearson1900x} to analyze the pairwise comparisons. The $\chi^2$ test evaluates whether observed preference selection frequencies deviate significantly from expected frequencies under the null hypothesis of no association between dataset size and user preference. We implement this analysis using SciPy~\cite{scipy}.

\subsection{Study Results}
\begin{figure*}[!ht]
    \centering
    \begin{subfigure}[b]{0.8\textwidth}
        \includegraphics[width=\textwidth]{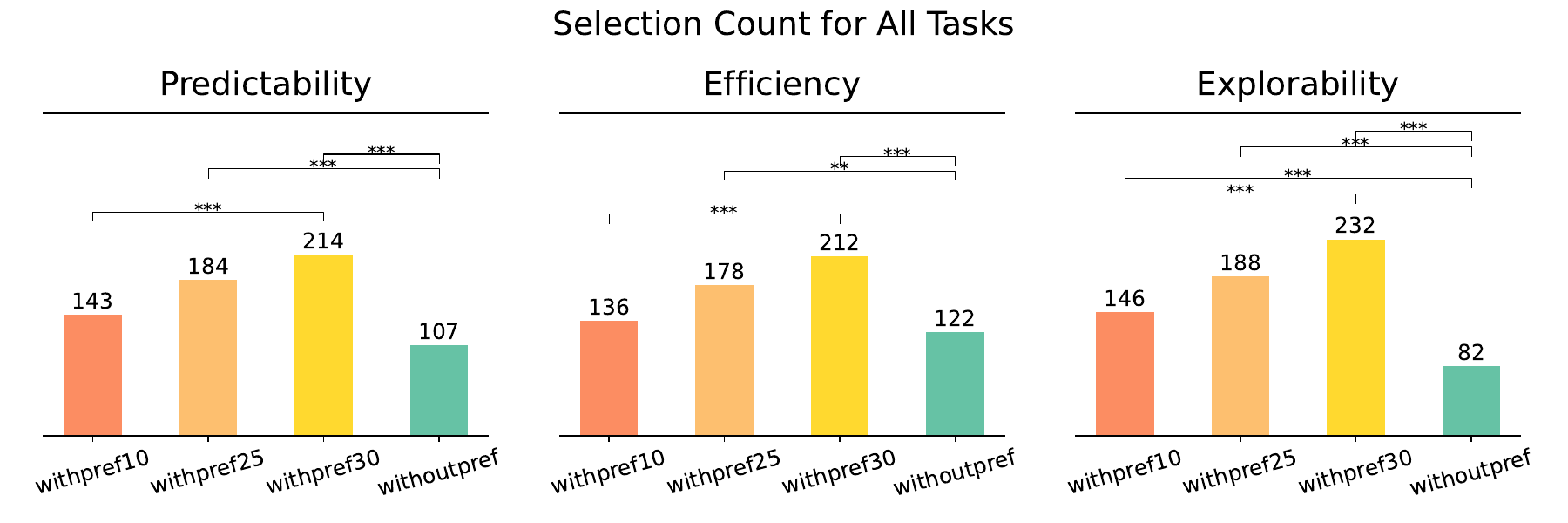}
        \caption{Total selection counts.}
        \label{fig:user_eval_per_aspect}
    \end{subfigure}
    \par\vspace{0.3cm}
    \begin{subfigure}[b]{0.8\textwidth}
        \includegraphics[width=\textwidth]{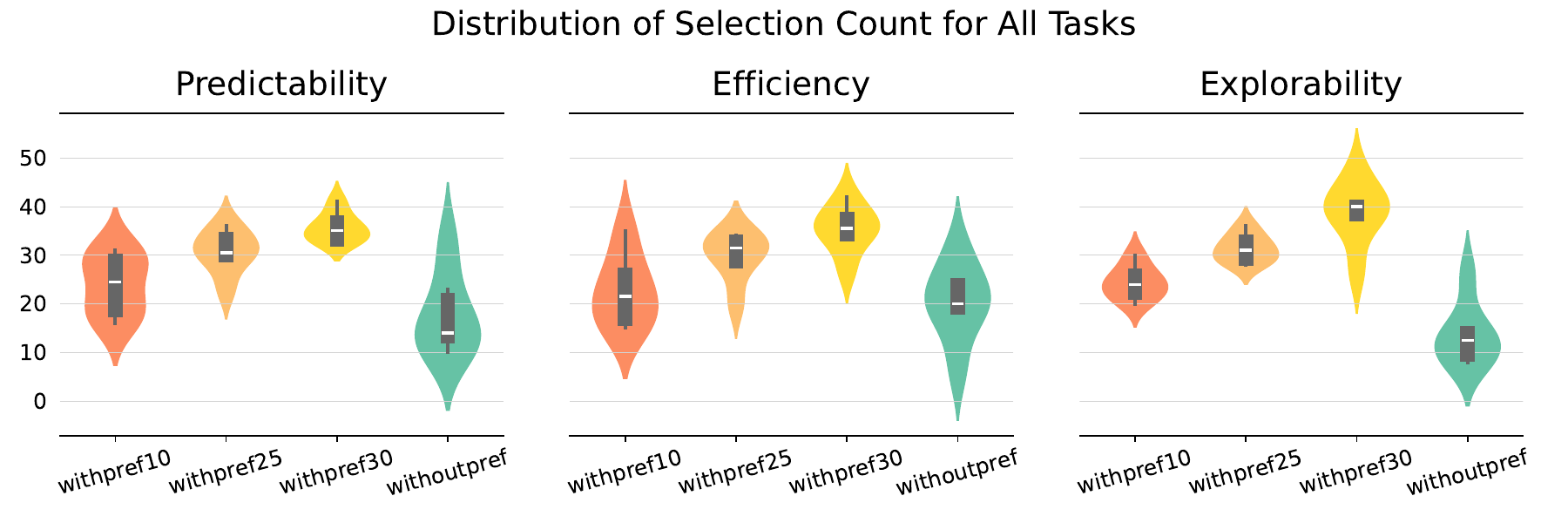}
        \caption{Distribution of total selection counts. The violin plots show the median, quartiles (25th and 75th percentiles), and the data range.}
        \label{fig:user_eval_per_aspect_violin}
    \end{subfigure}
    \caption{User study results across all tasks. Number and distribution of user preference selection counts of the three user preference aspects ($*$: $p < 0.05$, $**$: $p < 0.01$, $***$: $p < 0.001$).}
    \label{fig:user_eval_per_aspect_all}
    \Description{This figure contains two sub-figures: (a) shows the total selection counts as bar charts; (b) shows the distribution of total selection counts as violin plots. Both sub-figures consist of three columns that present the results of the three preference aspects.}
\end{figure*}

We present the user study results across all tasks in Figure~\ref{fig:user_eval_per_aspect_all}. For all three user preference aspects, the UI controls generated with the user preference dataset were preferred over those without. Regarding the size of the user preference dataset, \texttt{withpref30} was consistently favored compared to \texttt{withpref10} and \texttt{withpref25}.

\begin{figure*}[!ht]
    \centering
    \begin{subfigure}[b]{0.8\textwidth}
        \includegraphics[width=\textwidth]{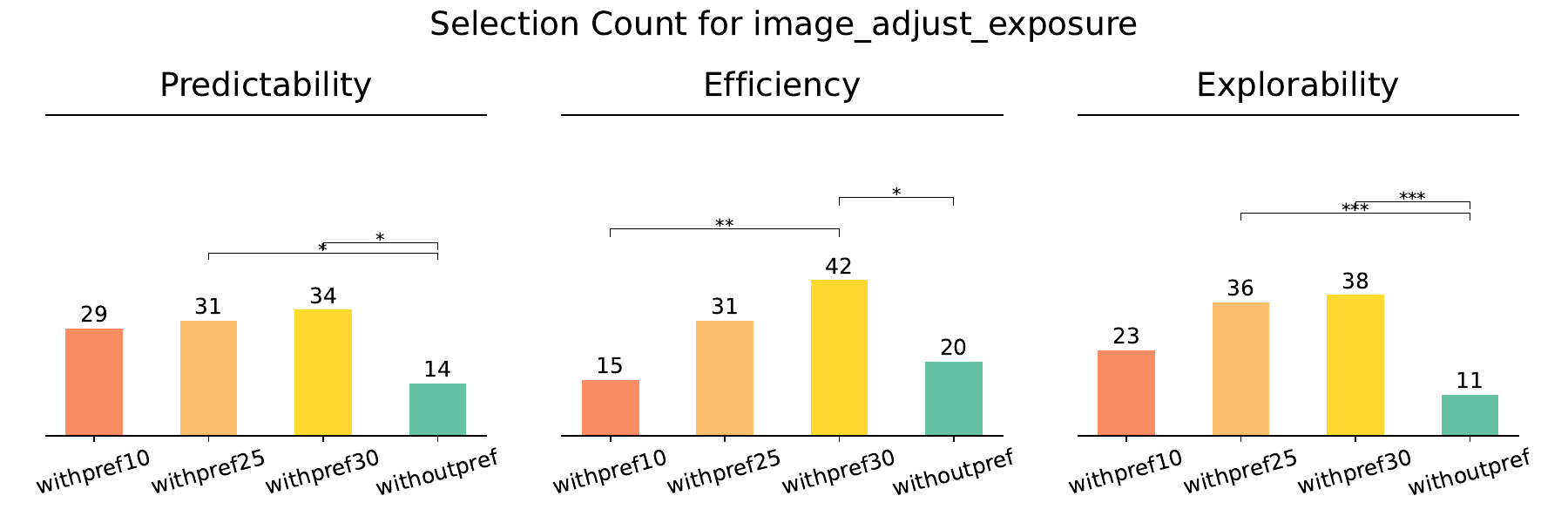}
    \end{subfigure}
    \par\vspace{0.3cm}
    \begin{subfigure}[b]{0.8\textwidth}
        \includegraphics[width=\textwidth]{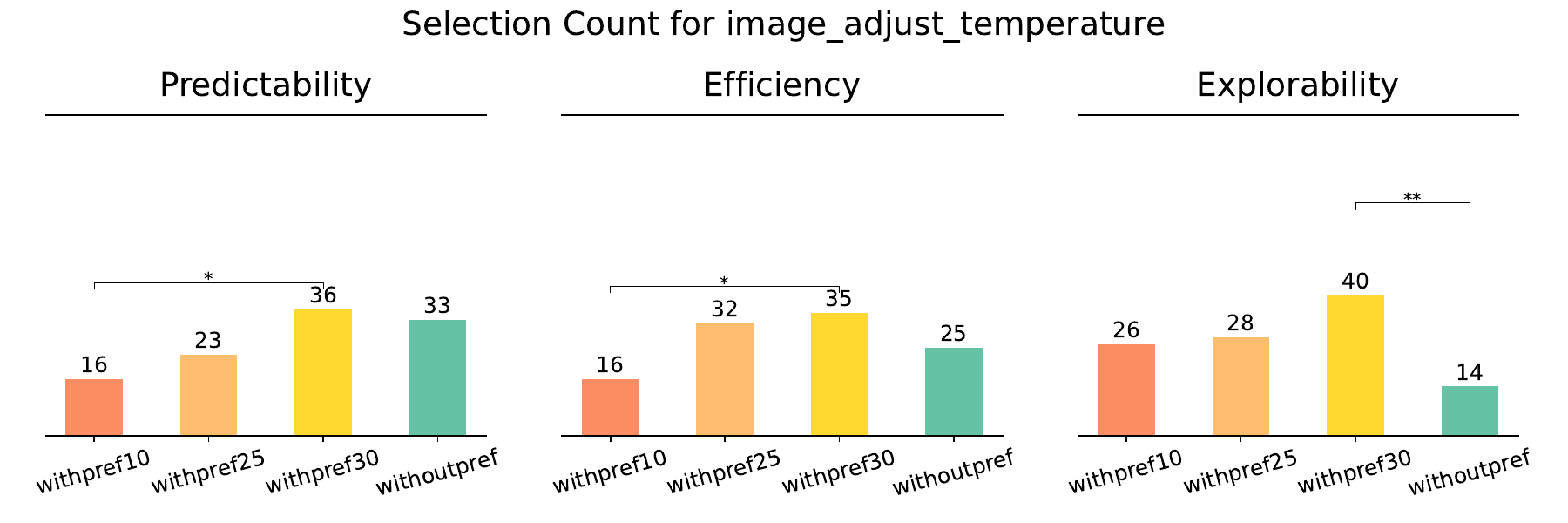}
    \end{subfigure}
    \par\vspace{0.3cm}
    \begin{subfigure}[b]{0.8\textwidth}
        \includegraphics[width=\textwidth]{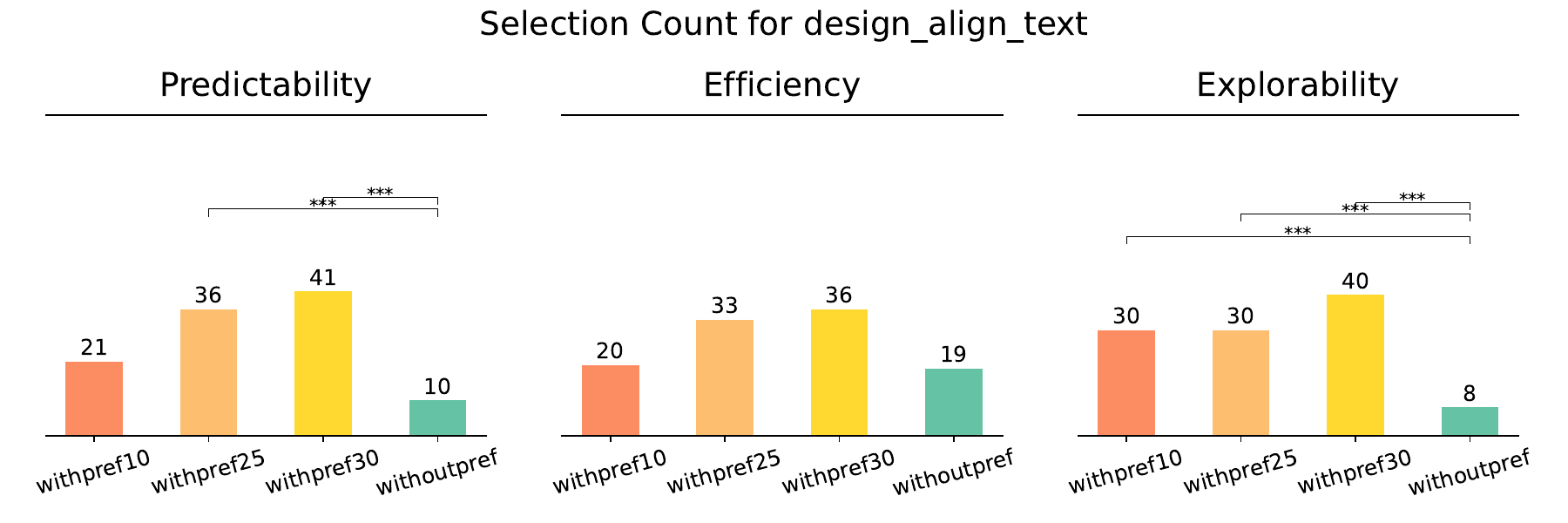}
    \end{subfigure}
    \caption{User study results of each task in task set 1. User preference selection counts of the three user preference aspects ($*$: $p < 0.05$, $**$: $p < 0.01$, $***$: $p < 0.001$).}
    \Description{This figure contains three sub-figures in three rows, showing the total selection counts as bar charts for tasks in task set 1. All sub-figures consist of three columns that present the results of the three preference aspects.}
    \label{fig:user_eval_per_task_set1}
\end{figure*}

\begin{figure*}[!ht]
    \centering
    \begin{subfigure}[b]{0.8\textwidth}
        \includegraphics[width=\textwidth]{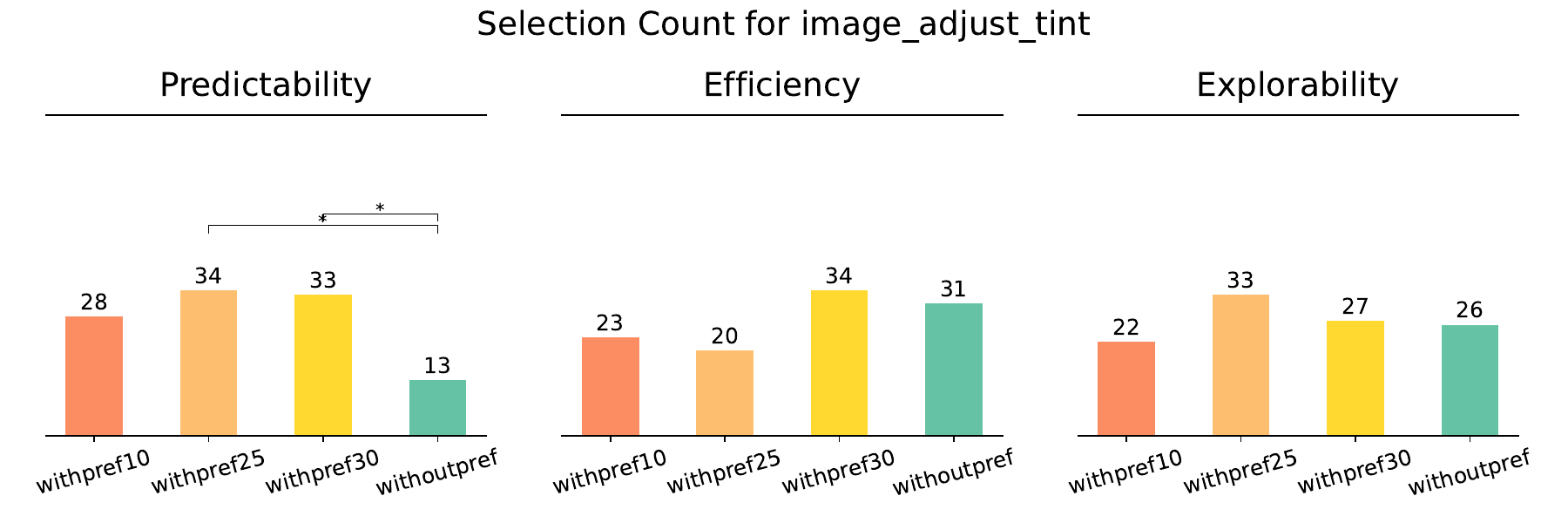}
    \end{subfigure}
    \par\vspace{0.3cm}
    \begin{subfigure}[b]{0.8\textwidth}
        \includegraphics[width=\textwidth]{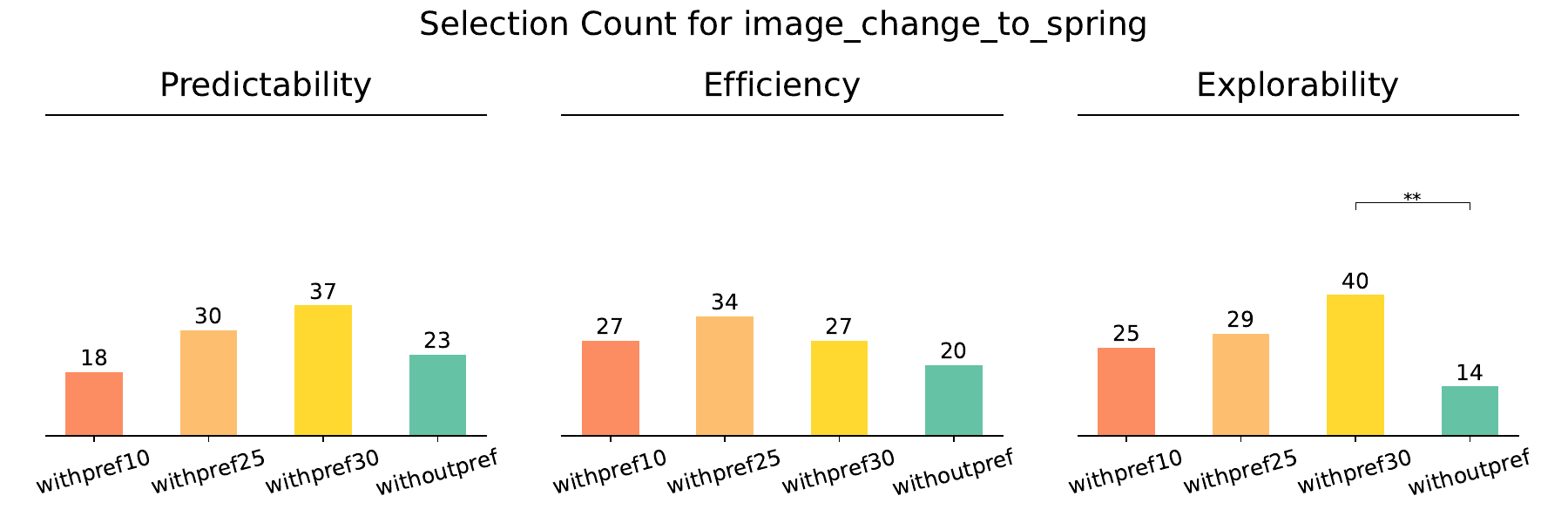}
    \end{subfigure}
    \par\vspace{0.3cm}
    \begin{subfigure}[b]{0.8\textwidth}
        \includegraphics[width=\textwidth]{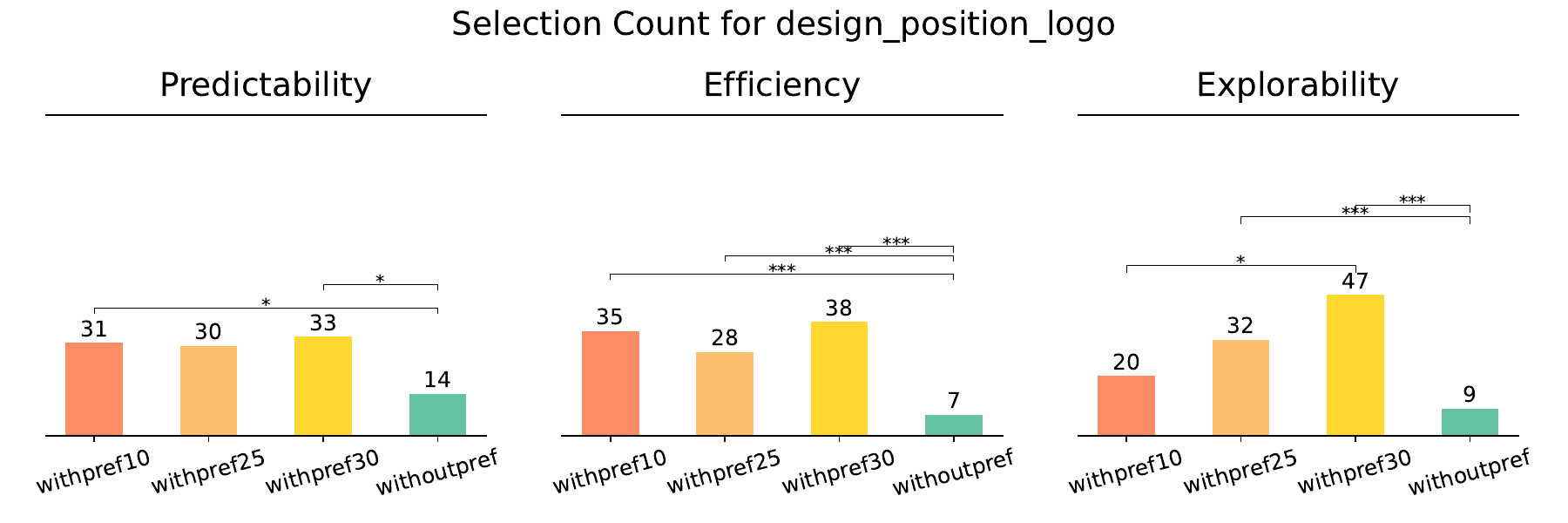}
    \end{subfigure}
    \caption{User study results of each task in task set 2. User preference selection counts of the three user preference aspects ($*$: $p < 0.05$, $**$: $p < 0.01$, $***$: $p < 0.001$).}
    \Description{This figure contains three sub-figures in three rows, showing the total selection counts as bar charts for tasks in task set 2. All sub-figures consist of three columns that present the results of the three preference aspects.}
    \label{fig:user_eval_per_task_set2}
\end{figure*}

Looking into each task (Figures~\ref{fig:user_eval_per_task_set1} and~\ref{fig:user_eval_per_task_set2}), for most tasks, \texttt{with\-pref30} and \texttt{withpref25} frequently converged as the most preferred configurations. For tasks like \texttt{image\_adjust\_exposure}, \texttt{de\-sign\_align\_text}, \texttt{image\_adjust\_tint}, and \texttt{design\_position\-\_logo}, the small dataset \texttt{withpref10}, though not the most favored, was still rated decently regarding some of the preference aspects.

Based on these results, we reflect on our research questions and summarize the main findings: 

\begin{itemize}[left=0pt]
    \item \textbf{Generalizability of UI generation method to new tasks (RQ1).} Our crowdsourced user preference dataset can effectively enhance LLM-based UI reasoning and generation, not only for non-overlapping image editing tasks but also for tasks beyond image editing (Tasks 5 and 6; text alignment and logo positioning).

    \item \textbf{Preference alignment of generated UIs guided by the user preference dataset (RQ2).} Integrating the user preference dataset enables the LLM to generate UI controls that more accurately align with target users' preferences regarding predictability, efficiency, and explorability, while its absence leads to generic UI controls.

    \item \textbf{Impact of dataset sizes on UI generation (RQ3).} The larger user preference datasets (\texttt{withpref30} and \texttt{withpref25}) allow the most consistently user preference-aligned UI control generation across tasks, while the dataset with fewer user preferences (\texttt{withpref10}) shows greater variability in user-rated preference alignment.
\end{itemize}

\section{Discussion} \label{sec:discussion}
In this section, we discuss our study findings and their implications on user-aligned UI design for broader tasks and user groups. 

\subsection{Supporting UI Generalizability with Curated Tasks}
\hspace*{1em} \textbf{Task curation for generalizable LLM-generated UIs.}
Our user study findings indicate that the user preference dataset helped to improve the LLM's reasoning to generate UIs for new and relevant tasks. For instance, a user preference dataset on image tone adjustment can indicate user preferences for broader tasks involving continuous color adjustment. Our work has demonstrated the generalizability of collecting user data on tasks that belong to adjusting continuous and discrete values, and requiring instant visual (color and position) feedback. This highlights the effectiveness of curated task design to guide task-generalizable UI reasoning and generation. 

\textbf{Breadth and depth of task domains.}
To enhance the user data's generalizability to guide the UI generation for broader tasks, extending the breadth and depth of task domains can be explored. On the one hand, UI interactions for different types of content edit can be considered, such as video, audio, and visualization. These task domains often involve distinct interaction requirements. For example, scrubbers or frame-level sliders for temporal control in video editing, up/down arrows or hand gestures for volume control in audio editing, and slider or knob for axis label rotation in data visualization. Due to these diverse interaction requirements, the UI design insights from one domain may not easily generalize to others. As such, multi-domain user data collection is needed to align LLM-based UI generation with niche interaction requirements and user preferences.

On the other hand, a deeper dive into specific task domains can help to derive in-depth UI designs assisted by user data. In image editing, for example, we can extend from single-value to multi-dimensional parameter adjustment. This is a common need as many image editing tasks consist of a combination of multiple parameters. For instance, to correct a low-exposure image, a user needs to adjust both image saturation and brightness. Although a common practice is to separately adjust the individual parameters, if the user looks for an efficient alternative, an UI control that intelligently combines multi-parameter adjustment may be desired. Accordingly, investigating extended task complexities can help uncover the dependence of parameters and the relevance of high-level task requirements, and augment the LLM's applicability to generate UIs for complicated task contexts. 

\subsection{Designing User-Centered UIs with Target User Data}
\hspace*{1em} \textbf{Gaps in UI design and target user preference.}
As presented in our formative study findings (Section~\ref{sec:prelim_study_results}) and crowdsourced user preference data (Section~\ref{sec:crowdsourcing_result_analysis}), user preferences for UI controls varied greatly when the task and preference aspect differed. These user-preferred UI controls also contrasted with the pre-defined UI controls in most UIs, such as software, website, and mobile, where designers make the call to decide which UI controls are used. Taking Photoshop, an image editing software tool, as an example. It offers a slider and text field for adjusting image lightness, saturation, and hue. However, our user data shows different UI controls that user preferred for their personalized needs (Table~\ref{tab:prelim_pref_10}). These revealed gaps emphasize the need to align UI designs with the niche preferences of the target users. 

\textbf{Target user data collection for user-aligned UI generation.}
To allow user-aligned UI design, our work has introduced a method to integrate user preferences into LLM-based UI generation. This method can evolve to include a broader range of UI design dimensions for multiple user populations. For instance, among established UI design principles~\cite{mckay2013ui, lidwell2010universal, chen2021should, shneiderman2000creating, norman1983design}, accessibility plays a crucial role for UI design and can benefit from user data collection from people with disabilities, e.g., the UI requirements and preferences of users with visual or motor impairments~\cite{gajos2004supple, gajos2007automatically}. By selecting the UI design principles of interest and collecting data from the users who value these principles, LLMs can be equipped with target user data to generate user-centered design insights. 

\subsection{Generating Personalized UIs with User Data}
\hspace*{1em} \textbf{Evolving user data for personalized UI generation.}
Based on comparing the impact of user preference dataset sizes, our study has found that larger datasets allowed more consistently user preference-aligned UI generation, while smaller datasets lost preference consistency but still performed decently on some tasks and user preference aspects. The smaller datasets are quick to collect, offering opportunities for personalized UI generation. For example, if a user is learning Photoshop to work on image editing tasks, a small dataset that reflects the user's preferences and needs can be collected to generate a beginner-friendly UI for their tasks to help them get started. As the user becomes familiar and confident to work on more advanced tasks, their dataset can continue to refresh or scale up to guide the evolution of generated UIs that match their expertise development. 

\textbf{UI personalization based on task workflows.}
With a similar idea, UIs can be personalized according to individual task workflows. Especially for expert users, personalized workflows are developed from years of experience and are often preferred to boost their productivity. UIs with UI elements that reflect their individual needs can become a catalyst to augment personalized workflows. Although existing UI design has enabled features like hiding user-selected tools to reduce redundancy for UI personalization~\cite{photoshop, blender}, the UIs and workflows of user tasks are largely independent. A personalized user dataset, however, can collect user-preferred tools and features, and guide the generation of UIs that select and organize tools of interest according to the workflows that users prefer. This transforms UI design into a way that UIs are tailored for users instead of requiring user learning and adaption of UIs.

\section{Limitations and Future Work} \label{sec:limitations}
In this section, we discuss the limitations of our work and offer insights for future research.

\textbf{New interaction modalities beyond desktop UIs.}
Although our work mainly focused on desktop UI generation, user-aligned UI design for other interaction modalities, such as UIs in AR/VR, wearable device UIs, and tangible UIs, is also critical. It would be interesting to extend our method to these contexts. Future work can collect user data regarding the user needs and preferences in various types of UI interaction and evaluate the user data's impact on UI design generation and user experience. 

\textbf{Multi-cultural user data collection.} 
We collected user data from English-speaking participants. This can overlook the cultural nuance of user preferences in broader global contexts. In UI design, user preferences for color schemes, layout conventions, and interaction metaphors may differ across linguistic and cultural divides~\cite{reinecke2011improving}. Future work could enhance cross-cultural inclusiveness in the data collection process, which would allow LLMs to generate UIs to align with users from different cultural backgrounds.

\textbf{LLMs for UI reasoning and generation.}
Our implementation used GPT-4o for UI reasoning and generation. To utilize the most of LLM capabilities, future research could test different LLMs for the reasoning and code generation. For example, models like GPT-5 or Gemini 3 can be used for advanced reasoning, Claude 4.5 Opus or GPT-Codex can be used for UI code generation, and open-source models like Gemma or DeepSeek can be used for lower cost and easier fine-tuning for specific task domains. This can also lead to an in-depth comparison of different LLMs' performance on user preference-aligned UI implementation.

\section{Conclusion} \label{sec:conclusion}
In this paper, we introduced the \crowdgenui{} method for user preference-aligned UI design with LLM-based UI generation. This method used a user preference dataset to direct the reasoning and implementation of generative UIs through enhanced LLM reasoning and code generation. To implement this method, we crowdsourced a user preference dataset with 720 pieces of user preferences from 50 general users, and used this dataset to guide the generation of UIs for multiple image editing tasks. Through a user study with 72 additional general users to evaluate the generated UIs, we demonstrated the method's generalizability to support UI reasoning for different but related tasks, effectiveness to align UI generation with user preferences, and scalability that allows consistent user preference alignment with larger datasets and cost-effective user preference capture with smaller datasets. We further discussed the user study results to explore how our method can be applied to support LLM-based UI design for broader task domains and user groups, as well as personalized user needs.




\bibliographystyle{ACM-Reference-Format}
\bibliography{bibliography}

@online{prolific,
  author = {Prolific},
  title = {Prolific},
  year = 2024,
  url = {https://www.prolific.com/},
  update = {2024-09},
  organization={Prolific}
}

@online{ipywidgets,
  author = {Jupyter},
  title = {Jupyter Widgets},
  year = 2024,
  url = {https://ipywidgets.readthedocs.io/en/stable/},
  update = {2024-09},
  organization={Jupyter}
}

@online{scipy,
  author = {SciPy},
  title = {SciPy},
  year = 2024,
  url = {https://scipy.org/},
  update = {2024-10},
  organization={SciPy}
}

@online{photoshop,
  author = {Adobe},
  title = {Adobe Photoshop},
  year = 2024,
  url = {https://www.adobe.com/products/photoshop.html},
  update = {2024-10},
  organization={Adobe}
}

@online{blender,
  author = {Blender},
  title = {Blender},
  year = 2024,
  url = {https://www.blender.org/},
  update = {2024-10},
  organization={Blender}
}

@online{gpt4o,
  author = {OpenAI},
  title = {OpenAI GPT-4o},
  year = 2024,
  url = {https://openai.com/index/hello-gpt-4o/},
  update = {2024-10},
  organization={OpenAI}
}

@book{mckay2013ui,
  title={UI is communication: How to design intuitive, user centered interfaces by focusing on effective communication},
  author={McKay, Everett N},
  year={2013},
  publisher={Newnes}
}

@inproceedings{vaithilingam2024dynavis,
  title={DynaVis: Dynamically Synthesized UI Widgets for Visualization Editing},
  author={Vaithilingam, Priyan and Glassman, Elena L and Inala, Jeevana Priya and Wang, Chenglong},
  booktitle={Proceedings of the CHI Conference on Human Factors in Computing Systems},
  pages={1--17},
  year={2024}
}

@article{cheng2024biscuit,
  title={BISCUIT: Scaffolding LLM-Generated Code with Ephemeral UIs in Computational Notebooks},
  author={Cheng, Ruijia and Barik, Titus and Leung, Alan and Hohman, Fred and Nichols, Jeffrey},
  journal={arXiv preprint arXiv:2404.07387},
  year={2024}
}

@book{lidwell2010universal,
  title={Universal principles of design, revised and updated: 125 ways to enhance usability, influence perception, increase appeal, make better design decisions, and teach through design},
  author={Lidwell, William and Holden, Kritina and Butler, Jill},
  year={2010},
  publisher={Rockport Pub}
}

@inproceedings{gajos2004supple,
  title={SUPPLE: automatically generating user interfaces},
  author={Gajos, Krzysztof and Weld, Daniel S},
  booktitle={Proceedings of the 9th international conference on Intelligent user interfaces},
  pages={93--100},
  year={2004}
}

@inproceedings{gajos2007automatically,
  title={Automatically generating user interfaces adapted to users' motor and vision capabilities},
  author={Gajos, Krzysztof Z and Wobbrock, Jacob O and Weld, Daniel S},
  booktitle={Proceedings of the 20th annual ACM symposium on User interface software and technology},
  pages={231--240},
  year={2007}
}

@article{pearson1900x,
  title={X. On the criterion that a given system of deviations from the probable in the case of a correlated system of variables is such that it can be reasonably supposed to have arisen from random sampling},
  author={Pearson, Karl},
  journal={The London, Edinburgh, and Dublin Philosophical Magazine and Journal of Science},
  volume={50},
  number={302},
  pages={157--175},
  year={1900},
  publisher={Taylor \& Francis}
}

@article{barke2023grounded,
  title={Grounded copilot: How programmers interact with code-generating models},
  author={Barke, Shraddha and James, Michael B and Polikarpova, Nadia},
  journal={Proceedings of the ACM on Programming Languages},
  volume={7},
  number={OOPSLA1},
  pages={85--111},
  year={2023},
  publisher={ACM New York, NY, USA}
}

@book{colbourn2010crc,
  title={CRC handbook of combinatorial designs},
  author={Colbourn, Charles J},
  year={2010},
  publisher={CRC press}
}

@article{chen2021should,
  title={How should i improve the ui of my app? a study of user reviews of popular apps in the google play},
  author={Chen, Qiuyuan and Chen, Chunyang and Hassan, Safwat and Xing, Zhengchang and Xia, Xin and Hassan, Ahmed E},
  journal={ACM Transactions on Software Engineering and Methodology (TOSEM)},
  volume={30},
  number={3},
  pages={1--38},
  year={2021},
  publisher={ACM New York, NY, USA}
}

@article{foong2017novice,
  title={Novice and expert sensemaking of crowdsourced design feedback},
  author={Foong, Eureka and Gergle, Darren and Gerber, Elizabeth M},
  journal={Proceedings of the ACM on Human-Computer Interaction},
  volume={1},
  number={CSCW},
  pages={1--18},
  year={2017},
  publisher={ACM New York, NY, USA}
}

@article{lee2018exploring,
  title={Exploring real-time collaboration in crowd-powered systems through a ui design tool},
  author={Lee, Sang Won and Krosnick, Rebecca and Park, Sun Young and Keelean, Brandon and Vaidya, Sach and O'Keefe, Stephanie D and Lasecki, Walter S},
  journal={Proceedings of the ACM on Human-Computer Interaction},
  volume={2},
  number={CSCW},
  pages={1--23},
  year={2018},
  publisher={ACM New York, NY, USA}
}

@inproceedings{luther2015structuring,
  title={Structuring, aggregating, and evaluating crowdsourced design critique},
  author={Luther, Kurt and Tolentino, Jari-Lee and Wu, Wei and Pavel, Amy and Bailey, Brian P and Agrawala, Maneesh and Hartmann, Bj{\"o}rn and Dow, Steven P},
  booktitle={Proceedings of the 18th ACM conference on computer supported cooperative work \& social computing},
  pages={473--485},
  year={2015}
}

@inproceedings{xu2014voyant,
  title={Voyant: generating structured feedback on visual designs using a crowd of non-experts},
  author={Xu, Anbang and Huang, Shih-Wen and Bailey, Brian},
  booktitle={Proceedings of the 17th ACM conference on Computer supported cooperative work \& social computing},
  pages={1433--1444},
  year={2014}
}

@inproceedings{xu2015classroom,
  title={A classroom study of using crowd feedback in the iterative design process},
  author={Xu, Anbang and Rao, Huaming and Dow, Steven P and Bailey, Brian P},
  booktitle={Proceedings of the 18th ACM conference on computer supported cooperative work \& social computing},
  pages={1637--1648},
  year={2015}
}

@inproceedings{yu2016encouraging,
  title={Encouraging “outside-the-box” thinking in crowd innovation through identifying domains of expertise},
  author={Yu, Lixiu and Kittur, Aniket and Kraut, Robert E},
  booktitle={Proceedings of the 19th ACM Conference on Computer-Supported Cooperative Work \& Social Computing},
  pages={1214--1222},
  year={2016}
}

@article{oppenlaender2020crowdui,
  title={CrowdUI: Supporting web design with the crowd},
  author={Oppenlaender, Jonas and Tiropanis, Thanassis and Hosio, Simo},
  journal={Proceedings of the ACM on Human-Computer Interaction},
  volume={4},
  number={EICS},
  pages={1--28},
  year={2020},
  publisher={ACM New York, NY, USA}
}

@article{kumar2025llm,
  title={LLM Post-Training: A Deep Dive into Reasoning Large Language Models},
  author={Kumar, Komal and Ashraf, Tajamul and Thawakar, Omkar and Anwer, Rao Muhammad and Cholakkal, Hisham and Shah, Mubarak and Yang, Ming-Hsuan and Torr, Phillip HS and Khan, Salman and Khan, Fahad Shahbaz},
  journal={arXiv preprint arXiv:2502.21321},
  year={2025}
}

@article{hossain2015crowdsourcing,
  title={Crowdsourcing: a comprehensive literature review},
  author={Hossain, Mokter and Kauranen, Ilkka},
  journal={Strategic Outsourcing: An International Journal},
  volume={8},
  number={1},
  pages={2--22},
  year={2015},
  publisher={Emerald Group Publishing Limited}
}

@article{tomczak2023over,
  title={What over 1,000,000 participants tell us about online research protocols},
  author={Tomczak, Johanna and Gordon, Andrew and Adams, Jamie and Pickering, Jade S and Hodges, Nick and Evershed, Jo K},
  journal={Frontiers in Human Neuroscience},
  volume={17},
  pages={1228365},
  year={2023},
  publisher={Frontiers Media SA}
}

@online{usertesting2025sample,
  author = {User Testing University},
  title = {Sample size recommendations},
  year = 2025,
  url = {https://help.usertesting.com/hc/en-us/articles/14820712486941-Sample-size-recommendations#:~:text=well%20and%20where%20they%20stumble},
  update = {2025-04},
  organization={User Testing University}
}

@online{userinterviews2025preference,
  author = {User Interviews},
  title = {Preference Testing},
  year = 2025,
  url = {https://www.userinterviews.com/ux-research-field-guide-chapter/preference-testing#:~:text=Recruiting%20participants%20for%20preference%20tests},
  update = {2025-04},
  organization={User Interviews}
}

@article{shneiderman2000creating,
  title={Creating creativity: user interfaces for supporting innovation},
  author={Shneiderman, Ben},
  journal={ACM Transactions on Computer-Human Interaction (TOCHI)},
  volume={7},
  number={1},
  pages={114--138},
  year={2000},
  publisher={ACM New York, NY, USA}
}

@inproceedings{norman1983design,
  title={Design principles for human-computer interfaces},
  author={Norman, Donald A},
  booktitle={Proceedings of the SIGCHI conference on Human Factors in Computing Systems},
  pages={1--10},
  year={1983}
}

@article{leviathangenerative,
  title={Generative UI: LLMs are Effective UI Generators},
  author={Leviathan, Yaniv and Kalman, Dani Valevski Matan and Lumen, Danny and Molad, Eyal Segalis Eyal and Pasternak, Shlomi and Natchu, Vishnu and Nygaard, Valerie and Matias, Srinivasan Cheenu Venkatachary James Manyika Yossi}
}

@article{liu2025auggen,
  title={AugGen: Augmenting Task-Based Learning in Professional Creative Software with LLM-Generated Scaffolded UIs},
  author={Liu, Yimeng and Sra, Misha},
  journal={arXiv preprint arXiv:2511.23379},
  year={2025}
}

@inproceedings{cao2025generative,
  title={Generative and Malleable User Interfaces with Generative and Evolving Task-Driven Data Model},
  author={Cao, Yining and Jiang, Peiling and Xia, Haijun},
  booktitle={Proceedings of the 2025 CHI Conference on Human Factors in Computing Systems},
  pages={1--20},
  year={2025}
}

@article{giacomin2014human,
  title={What is human centred design?},
  author={Giacomin, Joseph},
  journal={The design journal},
  volume={17},
  number={4},
  pages={606--623},
  year={2014},
  publisher={Taylor \& Francis}
}

@article{tractinsky2000beautiful,
  title={What is beautiful is usable},
  author={Tractinsky, Noam and Katz, Adi S and Ikar, Dror},
  journal={Interacting with computers},
  volume={13},
  number={2},
  pages={127--145},
  year={2000},
  publisher={Oxford University Press Oxford, UK}
}

@inproceedings{lu2022bridging,
  title={Bridging the Gap between UX Practitioners’ work practices and AI-enabled design support tools},
  author={Lu, Yuwen and Zhang, Chengzhi and Zhang, Iris and Li, Toby Jia-Jun},
  booktitle={CHI Conference on Human Factors in Computing Systems Extended Abstracts},
  pages={1--7},
  year={2022}
}

@book{shneiderman2010designing,
  title={Designing the user interface: strategies for effective human-computer interaction},
  author={Shneiderman, Ben},
  year={2010},
  publisher={Pearson Education India}
}

@article{yuan2024towards,
  title={Towards Human-AI Synergy in UI Design: Enhancing Multi-Agent Based UI Generation with Intent Clarification and Alignment},
  author={Yuan, Mingyue and Chen, Jieshan and Hu, Yongquan and Feng, Sidong and Xie, Mulong and Mohammadi, Gelareh and Xing, Zhenchang and Quigley, Aaron},
  journal={arXiv preprint arXiv:2412.20071},
  year={2024}
}

@online{rabi2025designing,
  author = {Rabi},
  title = {Designing Human-in-the-Loop AI Interfaces That Empower Users},
  year = 2025,
  url = {https://www.thesys.dev/blogs/designing-human-in-the-loop-ai-interfaces-that-empower-users},
  organization={Thesys}
}

@article{zeng2023evaluating,
  title={Evaluating large language models at evaluating instruction following},
  author={Zeng, Zhiyuan and Yu, Jiatong and Gao, Tianyu and Meng, Yu and Goyal, Tanya and Chen, Danqi},
  journal={arXiv preprint arXiv:2310.07641},
  year={2023}
}

@article{myers1995user,
  title={User interface software tools},
  author={Myers, Brad A},
  journal={ACM Transactions on Computer-Human Interaction (TOCHI)},
  volume={2},
  number={1},
  pages={64--103},
  year={1995},
  publisher={ACM New York, NY, USA}
}

@inproceedings{huang2019swire,
  title={Swire: Sketch-based user interface retrieval},
  author={Huang, Forrest and Canny, John F and Nichols, Jeffrey},
  booktitle={Proceedings of the 2019 CHI Conference on Human Factors in Computing Systems},
  pages={1--10},
  year={2019}
}

@inproceedings{deka2017rico,
  title={Rico: A mobile app dataset for building data-driven design applications},
  author={Deka, Biplab and Huang, Zifeng and Franzen, Chad and Hibschman, Joshua and Afergan, Daniel and Li, Yang and Nichols, Jeffrey and Kumar, Ranjitha},
  booktitle={Proceedings of the 30th annual ACM symposium on user interface software and technology},
  pages={845--854},
  year={2017}
}

@article{li2019layoutgan,
  title={Layoutgan: Generating graphic layouts with wireframe discriminators},
  author={Li, Jianan and Yang, Jimei and Hertzmann, Aaron and Zhang, Jianming and Xu, Tingfa},
  journal={arXiv preprint arXiv:1901.06767},
  year={2019}
}

@article{moran2018machine,
  title={Machine learning-based prototyping of graphical user interfaces for mobile apps},
  author={Moran, Kevin and Bernal-C{\'a}rdenas, Carlos and Curcio, Michael and Bonett, Richard and Poshyvanyk, Denys},
  journal={IEEE transactions on software engineering},
  volume={46},
  number={2},
  pages={196--221},
  year={2018},
  publisher={IEEE}
}

@inproceedings{wang2021screen2words,
  title={Screen2words: Automatic mobile ui summarization with multimodal learning},
  author={Wang, Bryan and Li, Gang and Zhou, Xin and Chen, Zhourong and Grossman, Tovi and Li, Yang},
  booktitle={The 34th Annual ACM Symposium on User Interface Software and Technology},
  pages={498--510},
  year={2021}
}

@article{myers2000past,
  title={Past, present, and future of user interface software tools},
  author={Myers, Brad and Hudson, Scott E and Pausch, Randy},
  journal={ACM Transactions on Computer-Human Interaction (TOCHI)},
  volume={7},
  number={1},
  pages={3--28},
  year={2000},
  publisher={ACM New York, NY, USA}
}

@inproceedings{subramonyam2021protoai,
  title={Protoai: Model-informed prototyping for ai-powered interfaces},
  author={Subramonyam, Hariharan and Seifert, Colleen and Adar, Eytan},
  booktitle={Proceedings of the 26th International Conference on Intelligent User Interfaces},
  pages={48--58},
  year={2021}
}

@online{figma,
  author = {Figma},
  title = {Figma},
  year = 2025,
  url = {https://www.figma.com},
  organization={Figma}
}

@online{sketch,
  author = {Sketch},
  title = {Sketch},
  year = 2025,
  url = {https://www.sketch.com},
  organization={Sketch}
}

@inproceedings{yuan2025understanding,
  title={Understanding and mitigating numerical sources of nondeterminism in llm inference},
  author={Yuan, Jiayi and Li, Hao and Ding, Xinheng and Xie, Wenya and Li, Yu-Jhe and Zhao, Wentian and Wan, Kun and Shi, Jing and Hu, Xia and Liu, Zirui},
  booktitle={The Thirty-ninth Annual Conference on Neural Information Processing Systems},
  year={2025}
}

@article{ouyang2022training,
  title={Training language models to follow instructions with human feedback},
  author={Ouyang, Long and Wu, Jeffrey and Jiang, Xu and Almeida, Diogo and Wainwright, Carroll and Mishkin, Pamela and Zhang, Chong and Agarwal, Sandhini and Slama, Katarina and Ray, Alex and others},
  journal={Advances in neural information processing systems},
  volume={35},
  pages={27730--27744},
  year={2022}
}

@article{rafailov2023direct,
  title={Direct preference optimization: Your language model is secretly a reward model},
  author={Rafailov, Rafael and Sharma, Archit and Mitchell, Eric and Manning, Christopher D and Ermon, Stefano and Finn, Chelsea},
  journal={Advances in neural information processing systems},
  volume={36},
  pages={53728--53741},
  year={2023}
}

@inproceedings{lu2025misty,
  title={Misty: Ui prototyping through interactive conceptual blending},
  author={Lu, Yuwen and Leung, Alan and Swearngin, Amanda and Nichols, Jeffrey and Barik, Titus},
  booktitle={Proceedings of the 2025 CHI Conference on Human Factors in Computing Systems},
  pages={1--17},
  year={2025}
}

@inproceedings{chen2025genui,
  title={The GenUI Study: Exploring the Design of Generative UI Tools to Support UX Practitioners and Beyond},
  author={Chen, Xiang'Anthony and Knearem, Tiffany and Li, Yang},
  booktitle={Proceedings of the 2025 ACM Designing Interactive Systems Conference},
  pages={1179--1196},
  year={2025}
}

@inproceedings{park2025leveraging,
  title={Leveraging Multimodal LLM for Inspirational User Interface Search},
  author={Park, Seokhyeon and Song, Yumin and Lee, Soohyun and Kim, Jaeyoung and Seo, Jinwook},
  booktitle={Proceedings of the 2025 CHI Conference on Human Factors in Computing Systems},
  pages={1--22},
  year={2025}
}

@article{lewis2020retrieval,
  title={Retrieval-augmented generation for knowledge-intensive nlp tasks},
  author={Lewis, Patrick and Perez, Ethan and Piktus, Aleksandra and Petroni, Fabio and Karpukhin, Vladimir and Goyal, Naman and K{\"u}ttler, Heinrich and Lewis, Mike and Yih, Wen-tau and Rockt{\"a}schel, Tim and others},
  journal={Advances in neural information processing systems},
  volume={33},
  pages={9459--9474},
  year={2020}
}

@article{wang2023aligning,
  title={Aligning large language models with human: A survey},
  author={Wang, Yufei and Zhong, Wanjun and Li, Liangyou and Mi, Fei and Zeng, Xingshan and Huang, Wenyong and Shang, Lifeng and Jiang, Xin and Liu, Qun},
  journal={arXiv preprint arXiv:2307.12966},
  year={2023}
}

@book{norman1988psychology,
  title={The psychology of everyday things.},
  author={Norman, Donald A},
  year={1988},
  publisher={Basic books}
}

@book{nielsen1994usability,
  title={Usability engineering},
  author={Nielsen, Jakob},
  year={1994},
  publisher={Morgan Kaufmann}
}

@article{mcinerney2000ui,
  title={The UI design process},
  author={McInerney, Paul and Sobiesiak, Rick},
  journal={ACM SIGCHI Bulletin},
  volume={32},
  number={1},
  pages={17--21},
  year={2000},
  publisher={ACM New York, NY, USA}
}

@article{olsson2004active,
  title={What active users and designers contribute in the design process},
  author={Olsson, Eva},
  journal={Interacting with computers},
  volume={16},
  number={2},
  pages={377--401},
  year={2004},
  publisher={Oxford University Press Oxford, UK}
}

@article{gould1985designing,
  title={Designing for usability: key principles and what designers think},
  author={Gould, John D and Lewis, Clayton},
  journal={Communications of the ACM},
  volume={28},
  number={3},
  pages={300--311},
  year={1985},
  publisher={ACM New York, NY, USA}
}

@article{zhou2023instruction,
  title={Instruction-following evaluation for large language models},
  author={Zhou, Jeffrey and Lu, Tianjian and Mishra, Swaroop and Brahma, Siddhartha and Basu, Sujoy and Luan, Yi and Zhou, Denny and Hou, Le},
  journal={arXiv preprint arXiv:2311.07911},
  year={2023}
}

@article{shao2024deepseekmath,
  title={Deepseekmath: Pushing the limits of mathematical reasoning in open language models},
  author={Shao, Zhihong and Wang, Peiyi and Zhu, Qihao and Xu, Runxin and Song, Junxiao and Bi, Xiao and Zhang, Haowei and Zhang, Mingchuan and Li, YK and Wu, Yang and others},
  journal={arXiv preprint arXiv:2402.03300},
  year={2024}
}

@article{reinecke2011improving,
  title={Improving performance, perceived usability, and aesthetics with culturally adaptive user interfaces},
  author={Reinecke, Katharina and Bernstein, Abraham},
  journal={ACM Transactions on Computer-Human Interaction (TOCHI)},
  volume={18},
  number={2},
  pages={1--29},
  year={2011},
  publisher={ACM New York, NY, USA}
}

@String{Computing = "Computing" }

@String{Computer = "{IEEE} Computer" }

@String{Chelsea = "Chelsea" }

\appendix

\newpage
\onecolumn

\section{LLM-Reasoned UI Controls for User Study Tasks} \label{appendix:widgets_of_user_evaluation_tasks}
Figures \ref{fig:reasoning_image_adjust_exposure} to \ref{fig:reasoning_design_position_logo} show the LLM-reasoned UI controls with crowdsourced datasets and with no dataset for the user study tasks.

\begin{figure*}[!ht]
    \centering
    \includegraphics[width=0.85\textwidth]{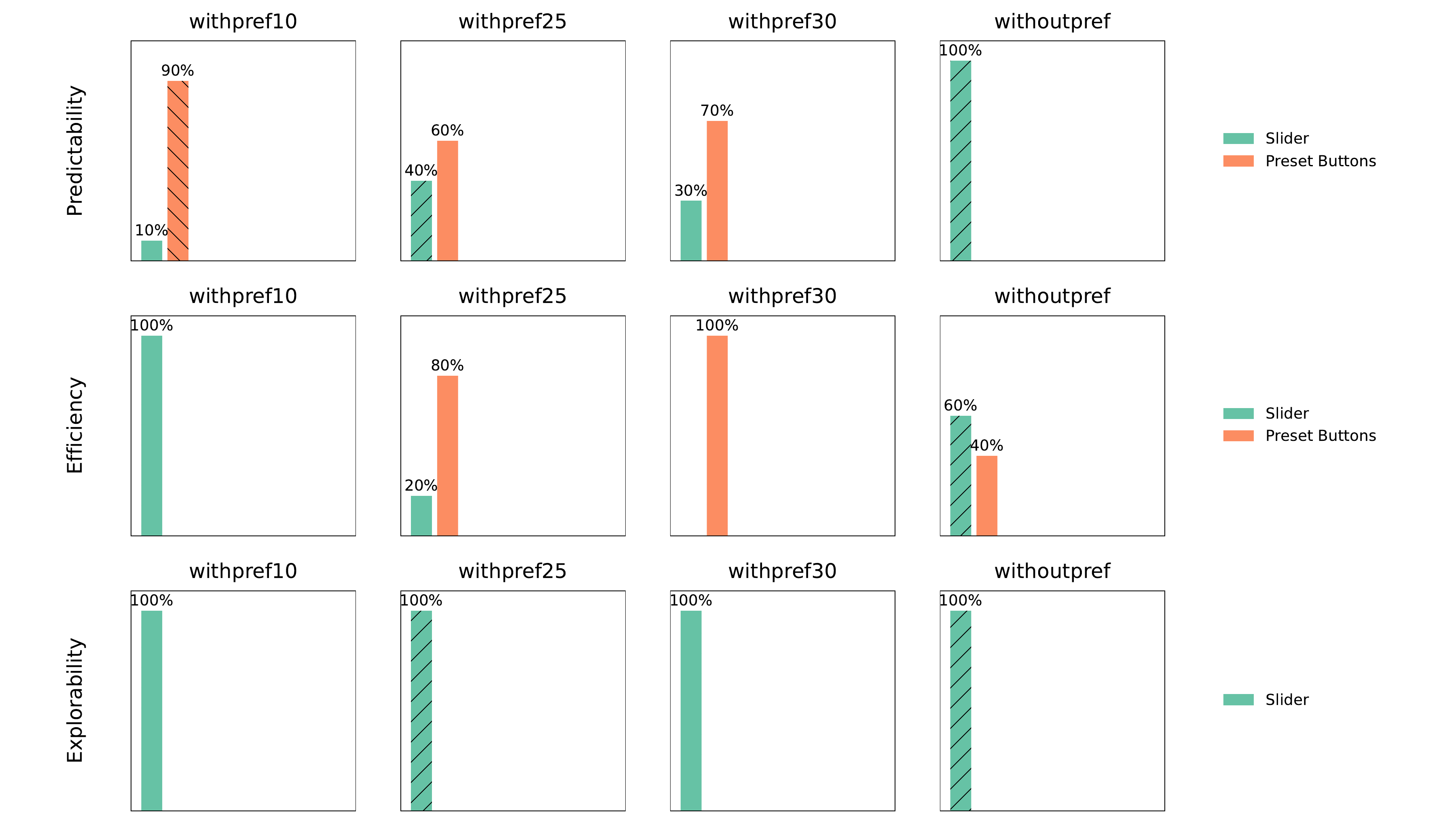}
    \caption{LLM-reasoned UI controls for task \texttt{image\_adjust\_exposure} with three sizes of the crowdsourced datasets (\texttt{withpref10}, \texttt{withpref25}, \texttt{withpref30}) and with no dataset (\texttt{withoutpref}).}
    \label{fig:reasoning_image_adjust_exposure}
\end{figure*}

\begin{figure*}[!ht]
    \centering
    \includegraphics[width=0.85\textwidth]{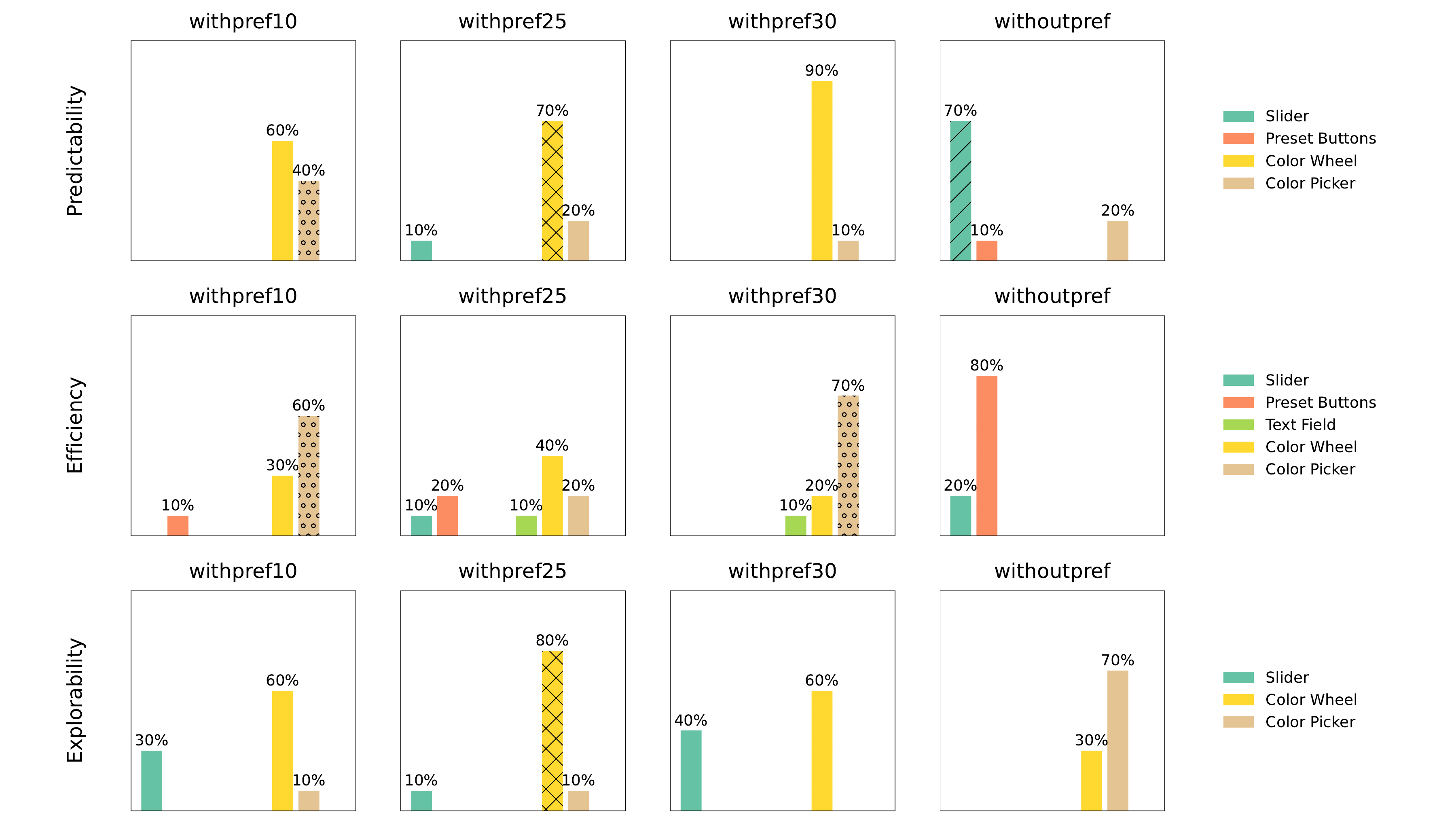}
    \caption{LLM-reasoned UI controls for task \texttt{image\_adjust\_tint} with three sizes of the crowdsourced datasets (\texttt{withpref10}, \texttt{withpref25}, \texttt{withpref30}) and with no dataset (\texttt{withoutpref}).}
    \label{fig:reasoning_image_adjust_tint}
\end{figure*}

\begin{figure*}[!ht]
    \centering
    \includegraphics[width=0.85\textwidth]{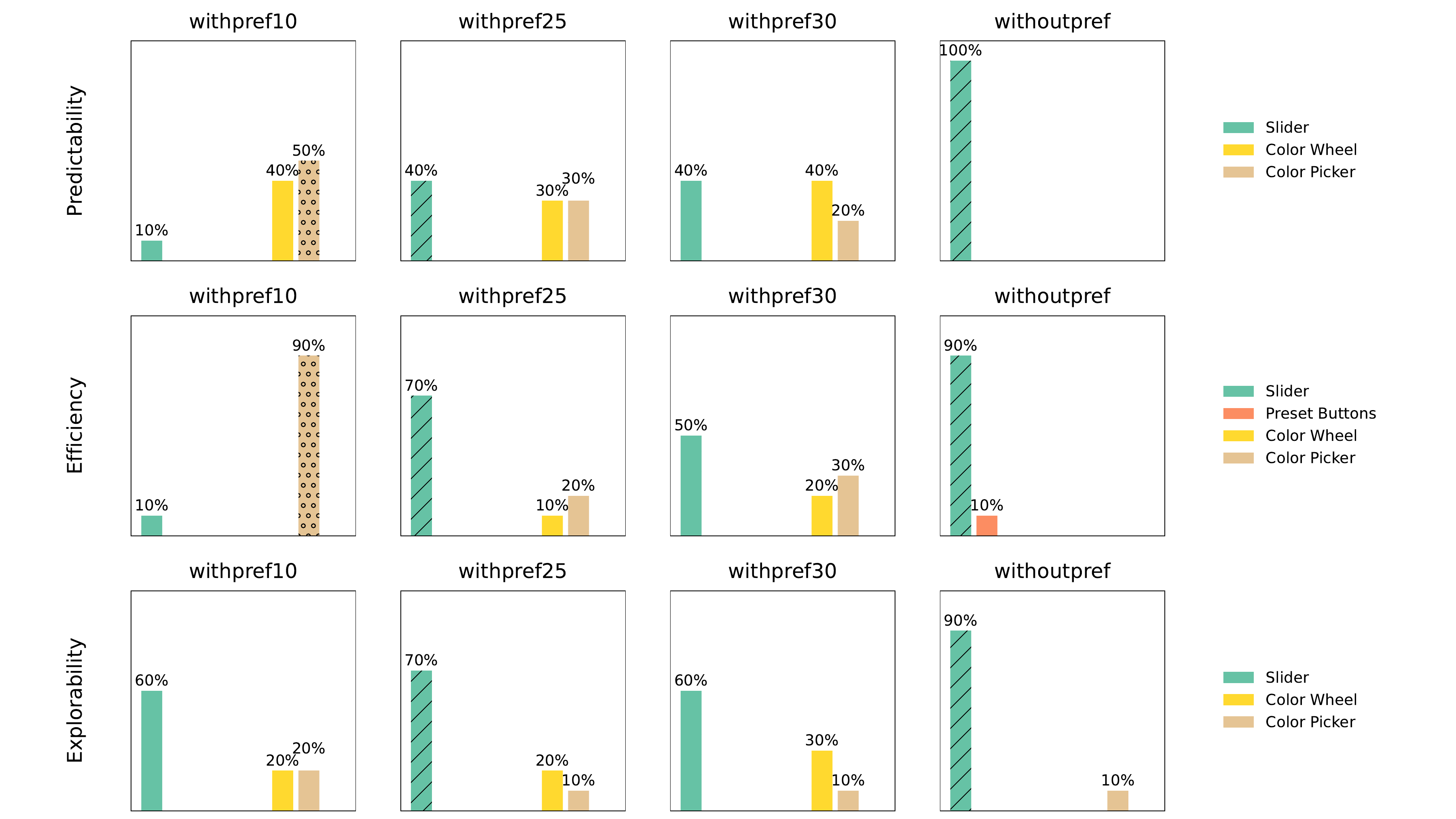}
    \caption{LLM-reasoned UI controls for task \texttt{image\_adjust\_temperature} with three sizes of the crowdsourced datasets (\texttt{withpref10}, \texttt{withpref25}, \texttt{withpref30}) and with no dataset (\texttt{withoutpref}).}
    \label{fig:reasoning_image_adjust_temperature}
\end{figure*}

\begin{figure*}[!ht]
    \centering
    \includegraphics[width=0.85\textwidth]{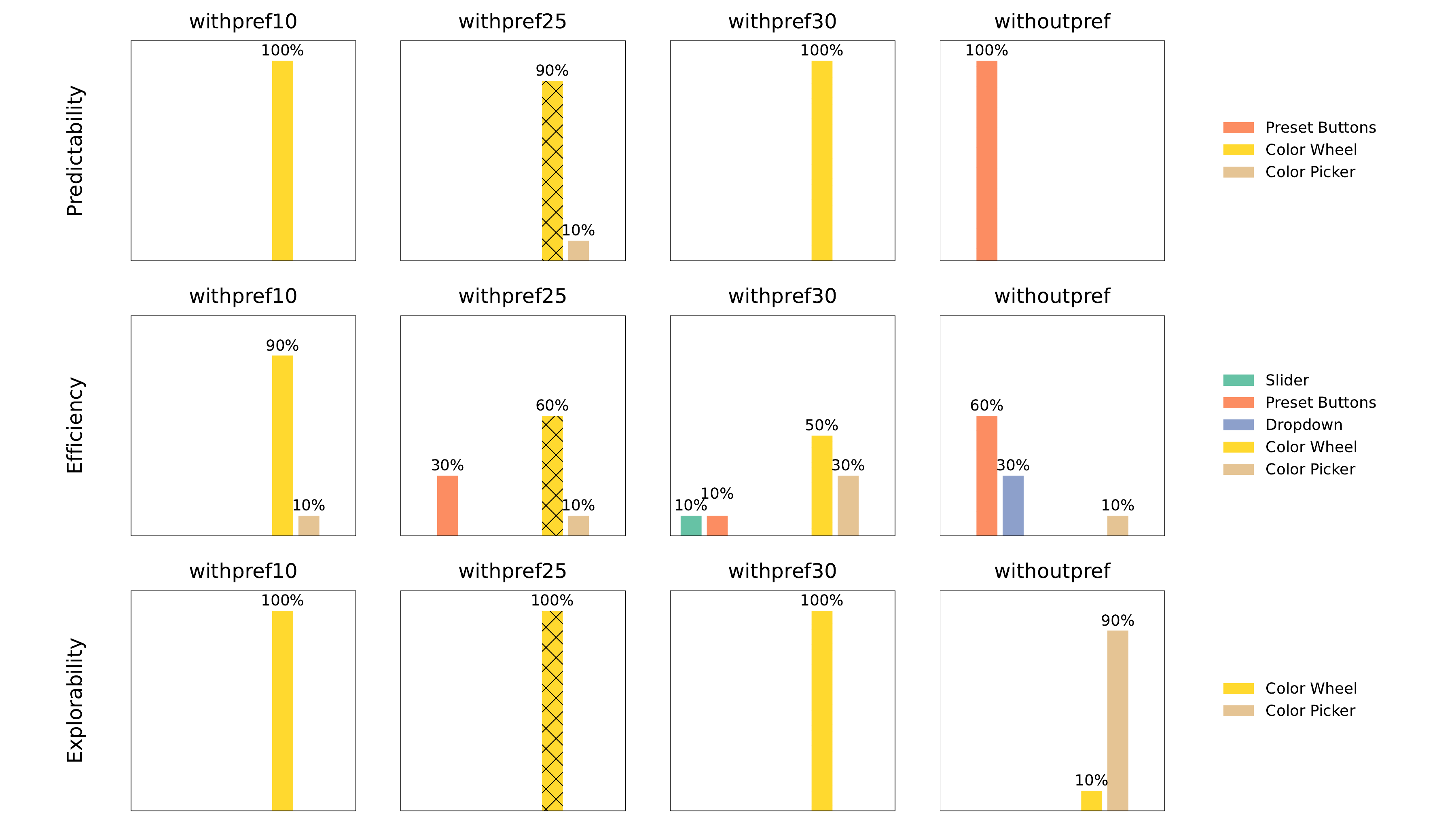}
    \caption{LLM-reasoned UI controls for task \texttt{image\_change\_to\_spring} with three sizes of the crowdsourced datasets (\texttt{withpref10}, \texttt{withpref25}, \texttt{withpref30}) and with no dataset (\texttt{withoutpref}).}
    \label{fig:reasoning_image_change_to_spring}
\end{figure*}

\begin{figure*}[!ht]
    \centering
    \includegraphics[width=0.85\textwidth]{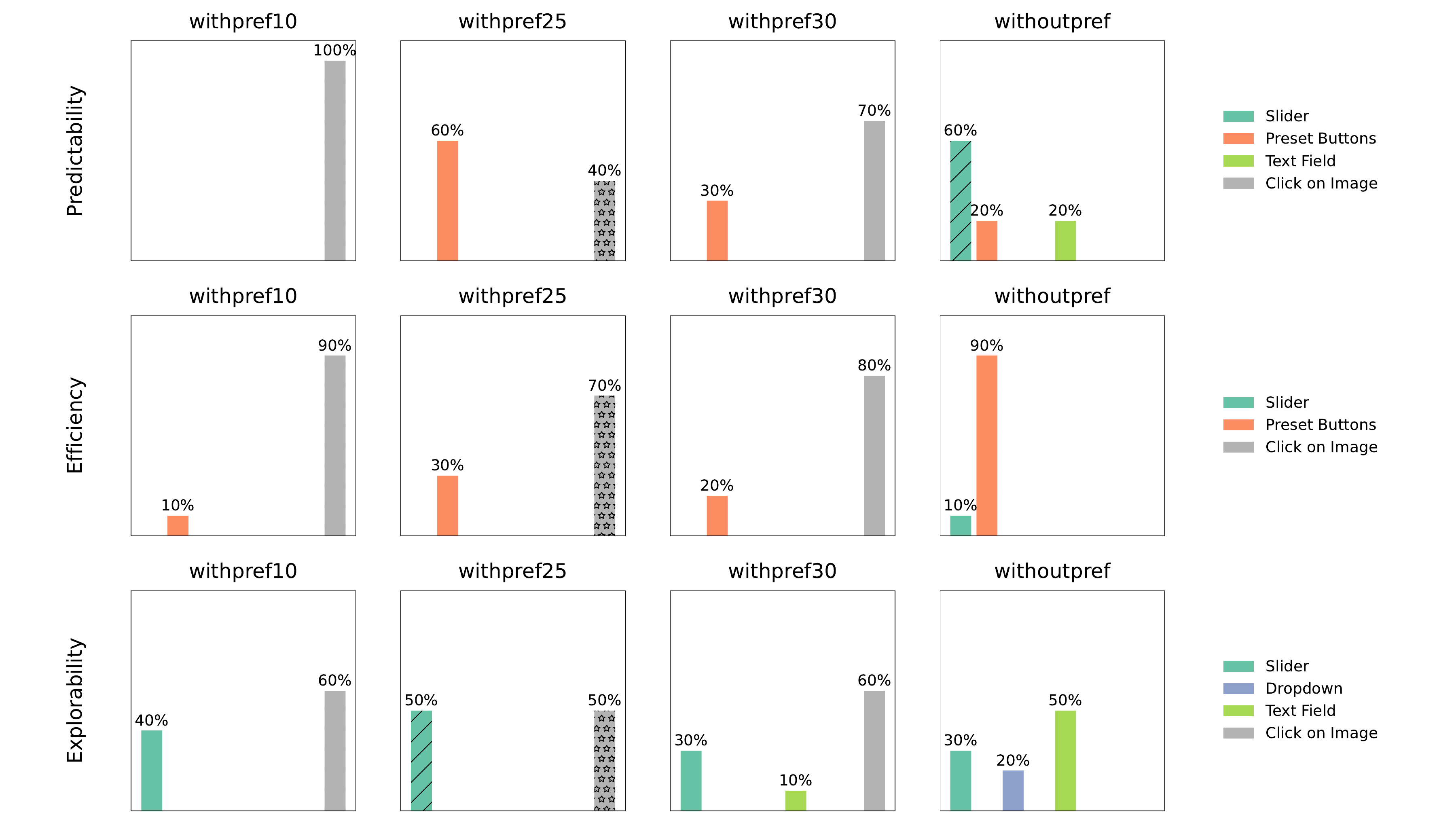}
    \caption{LLM-reasoned UI controls for task \texttt{design\_align\_text} with three sizes of the crowdsourced datasets (\texttt{withpref10}, \texttt{withpref25}, \texttt{withpref30}) and with no dataset (\texttt{withoutpref}).}
    \label{fig:reasoning_design_align_text}
\end{figure*}

\begin{figure*}[!ht]
    \centering
    \includegraphics[width=0.85\textwidth]{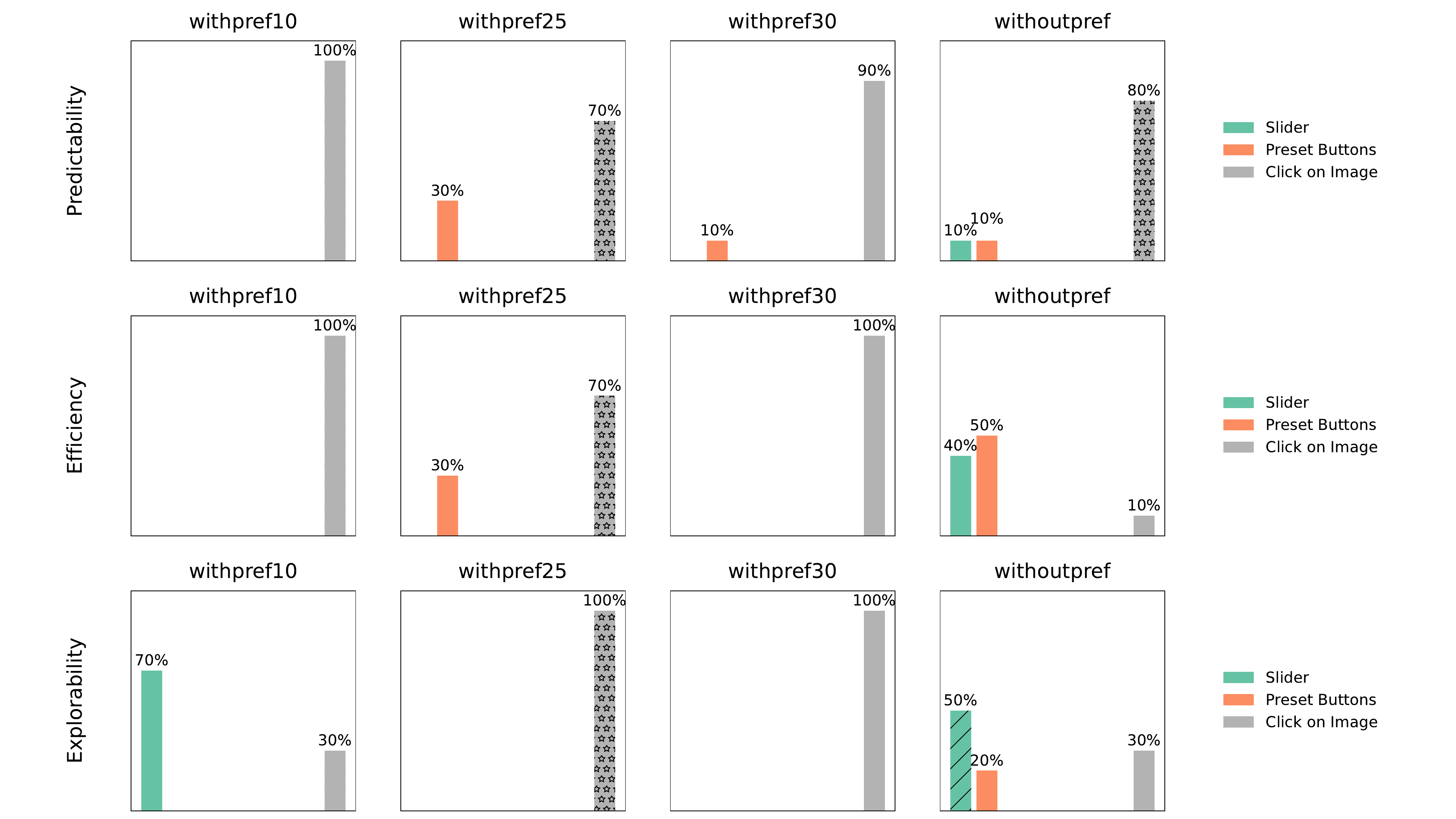}
    \caption{LLM-reasoned UI controls for task \texttt{design\_position\_logo} with three sizes of the crowdsourced datasets (\texttt{withpref10}, \texttt{withpref25}, \texttt{withpref30}) and with no dataset (\texttt{withoutpref}).}
    \label{fig:reasoning_design_position_logo}
\end{figure*}

\clearpage

\section{Prompts for LLM UI Control Reasoning and Code Generation}

\subsection{Prompts for UI Control Reasoning}
\label{appendix:prompt_widget_reasoning_withlib}
\begin{lstlisting}[language=json]
You are expert at reasoning UI controls that align with the user task and preference. You should follow the steps below for the reasoning.

First, please take the definitions of user preference aspects below:
    - Predictability: allows users to obtain results with no surprises.
    - Efficiency: allows users to perform tasks with a minimum amount of time and effort.
    - Explorability: allows users to explore multiple possibilities to perform the task.

Second, information of a crowdsourced UI control preference dataset is in the prompt:
    - Task description: detailed descriptions of all the tasks.
    - UI control preference aspects: the user preference aspects defined in user preference aspects.
        - UI control preference frequency: the frequency of user-preferred UI controls. Large numbers mean the corresponding UI control is preferred by more people. 
        - UI control preference reasons: the reasons for user-preferred UI controls.

Third, search for the most relevant tasks from the crowdsourced UI control preference dataset. 
    - You should compare the given task and the task descriptions in the dataset to help you to find the relevant tasks.
    - Focus on fundamental parameter adjustment requirements, e.g., tasks that require continuous value adjustment are similar, while tasks that require continuous or discrete value adjustment are not similar. 

Fourth, reason the most appropriate UI control for predictability, efficiency, and explorability. 
    - Your reasoning should be based on the relevant tasks you found in the dataset.
    - You must refer to the UI control preference reasons to assist your reasoning.
    - The UI control you reason must come from the given UI control candidates.

Lastly, based on your reasoning, organize your response in JSON format. Refer to the example below. 
{
    "user task": "<description of the provided user task>",
    "user preference aspect": "<description of the user-specified preference aspect>",
    "relevant tasks from the dataset": "<your reasoning>",
    "predictability_reasoning": {
        "<UI control type>": "<your reasoning>"
    },
    "efficiency_reasoning": {
        "<UI control type>": "<your reasoning>"
    },
    "explorability_reasoning": {
        "<UI control type>": "<your reasoning>"
    }
}
\end{lstlisting}

\subsection{Prompts for UI Control Code Generation}
\label{appendix:prompt_widget_coding}
\begin{lstlisting}[language=json]
You are expert at generating code for the offered UI controls to perform the specified task. You should follow the steps below for the code generation.

First, the UI control code you provide must allow users to perform the task specified in the content of user task.

Second, the UI control code you provide must refer to the example code for the implementation.

Third, organize your response in JSON format following the example responses below. Replace the content in "task", "control_type", and "control_code" with the actual information.
{
    "task": "Adjust image hue",
    "control_type": "Slider, Dropdown, Radio Buttons, Text Field, Preset Buttons, Color Wheel, Color Picker",
    "control_code": {
\end{lstlisting}
\begin{lstlisting}[language=Python]
import numpy as np
import matplotlib.pyplot as plt
import ipycontrols as controls
from IPython.display import display, clear_output
from PIL import Image
from skimage import data, img_as_ubyte
from matplotlib.patches import Wedge
import matplotlib.colors as mcolors

image = data.astronaut()
image = Image.fromarray(img_as_ubyte(image))

# Function to allow performing the task
def adjust_hue(image, hue):
    img_hsv = image.convert('HSV')  
    np_img = np.array(img_hsv)  

    hue_shift = int(hue * 255)
    
    np_img = np_img.astype(np.int32)
    np_img[..., 0] = (np_img[..., 0] + hue_shift) % 256

    np_img = np.clip(np_img, 0, 255).astype(np.uint8)

    adjusted_img = Image.fromarray(np_img, mode='HSV').convert('RGB') 
    return adjusted_img

# Function to create controls
def create_hue_controls():
    # Slider
    slider_label = controls.Label(value='Slider:')
    slider = controls.FloatSlider(value=0.0, min=0.0, max=1.0, step=0.01)

    # Dropdown
    dropdown_label = controls.Label(value='Dropdown:')
    dropdown = controls.Dropdown(options=[0.0, 0.2, 0.4, 0.6, 0.8], value=0.0)

    # Radio Buttons
    radio_buttons_label = controls.Label(value='Radio Buttons:')
    radio_buttons = controls.RadioButtons(options=[0.0, 0.2, 0.4, 0.6, 0.8], value=0.0)

    # Text Field
    text_field_label = controls.Label(value='Text Field:')
    text_field = controls.BoundedFloatText(value=0.0, min=0.0, max=1.0, step=0.01)
    
    # Preset Buttons
    preset_label = controls.Label(value='Preset buttons:')
    
    preset_hues = [
        (0.0, 'red'), 
        (0.2, 'green'), 
        (0.4, 'cyan'), 
        (0.6, 'blue'), 
        (0.8, 'magenta')
    ]
    
    hue_mapping = {f"{hue}": hue for hue, color in preset_hues}
    
    preset_buttons = [
        controls.Button(
            description=f"{hue}",
            layout=controls.Layout(width='75px', height='30px'),
            style={'button_color': color}
        )
        for hue, color in preset_hues
    ]
    
    preset_buttons_box = controls.HBox(preset_buttons)
    
    # Color Wheel
    color_wheel_label = controls.Label(value='Color Wheel:')
    color_wheel = controls.Output()
    create_color_wheel(color_wheel)
    
    # Color Picker
    color_picker_label = controls.Label(value='Color Picker:')
    color_picker = controls.ColorPicker(concise=True, value='#ffffff', disabled=False)

    # Layout
    layout = controls.Layout(align_items='flex-start')
    spacer = controls.Box(value='', layout=controls.Layout(height='20px'))

    # Combine controls into a vertical box
    controls_box = controls.VBox([slider_label, slider, spacer, 
                                dropdown_label, dropdown, spacer, 
                                radio_buttons_label, radio_buttons, spacer, 
                                text_field_label, text_field, spacer, 
                                preset_label, preset_buttons_box, spacer, 
                                color_wheel_label, color_wheel, spacer, 
                                color_picker_label, color_picker, spacer], layout=layout)
    
    return controls_box, slider, dropdown, radio_buttons, text_field, preset_buttons, hue_mapping, color_wheel, color_picker

# Function to create and display the color wheel
def create_color_wheel(output):
    with output:
        clear_output(wait=True)
        
        fig, ax = plt.subplots(figsize=(1.5, 1.5))
        
        num_colors = 360
        theta = np.linspace(0, 2 * np.pi, num_colors, endpoint=False)
        colors = plt.cm.hsv(theta / (2 * np.pi))
        
        for i in range(num_colors):
            wedge = Wedge(center=(0, 0), r=1, theta1=(i * 360 / num_colors), 
                          theta2=((i + 1) * 360 / num_colors), color=colors[i], 
                          transform=ax.transData._b, clip_on=False)
            ax.add_patch(wedge)
        
        ax.set_aspect('equal')
        ax.set_xlim(-1.1, 1.1)
        ax.set_ylim(-1.1, 1.1)
        ax.axis('off')
        fig.canvas.mpl_connect('button_press_event', on_color_wheel_click)
        plt.show()

# Function to handle color wheel click
def on_color_wheel_click(event):
    if event.inaxes:
        x, y = event.xdata, event.ydata
        theta = np.arctan2(y, x) % (2 * np.pi)
        hue = theta / (2 * np.pi)
        hue = round(hue, 2) 

        update_plot(None, image, output, slider, dropdown, radio_buttons, text_field, preset_buttons, hue_mapping, hue=hue)

# Function to convert hex color to hue value
def hex_to_hue(hex_color):
    rgb = np.array([int(hex_color[i:i+2], 16) for i in (1, 3, 5)]) / 255.0
    hsv = mcolors.rgb_to_hsv(rgb.reshape(1, 1, 3))
    hue = hsv[0, 0, 0]
    return hue

# Function to update the image display based on control values
def update_plot(change, image, output, slider, dropdown, radio_buttons, text_field, preset_buttons, hue_mapping, hue=None):
    with output:
        clear_output(wait=True)

        if hue is None:
            if change and change['owner'] in preset_buttons:
                clicked_button = change['owner']
                hue = hue_mapping[clicked_button.description]
            elif change and change['owner'] == slider:
                hue = slider.value
            elif change and change['owner'] == dropdown:
                hue = dropdown.value
            elif change and change['owner'] == radio_buttons:
                hue = radio_buttons.value
            elif change and change['owner'] == text_field:
                hue = text_field.value
            elif change and change['owner'] == color_picker:
                hue = hex_to_hue(change['new'])
            else:
                hue = 0.0

        adjusted_image = adjust_hue(image, hue)
        plt.figure(figsize=(4, 4))
        plt.imshow(adjusted_image)
        plt.axis('off')
        plt.title(f'Hue Adjustment: {hue:.2f}')
        plt.show()

# Function to link controls to the update function
def link_controls_to_update(image, output, slider, dropdown, radio_buttons, text_field, preset_buttons, hue_mapping, color_picker):
    def callback(change):
        update_plot(change, image, output, slider, dropdown, radio_buttons, text_field, preset_buttons, hue_mapping)
    
    slider.observe(callback, names='value')
    dropdown.observe(callback, names='value')
    radio_buttons.observe(callback, names='value')
    text_field.observe(callback, names='value')
    color_picker.observe(callback, names='value')
    for btn in preset_buttons:
        btn.on_click(lambda btn: update_plot({'owner': btn}, image, output, slider, dropdown, radio_buttons, text_field, preset_buttons, hue_mapping))
\end{lstlisting}
\begin{lstlisting}[language=json]
}
\end{lstlisting}


\section{User Study Instructions} \label{appendix:user_eval_briefing_instructions}

Welcome! This study collects your preferences for different UI controls.
Your participation is voluntary. You are free to quit the study at any time. You can participate in this study only once.
All your responses (including demographic info) will be kept private and only used for research purposes.

Please follow the study instructions below. 

\begin{itemize}[leftmargin=*, labelsep=1em]
    \item You will be working on \textbf{3 tasks}. Each task is followed by \textbf{6 questions} asking for your UI control preference by selecting one of two provided options for each question.

    \item For each task, please follow the task description and interact with \textbf{all the provided UI controls}.

    \item After interacting with the UI controls, you will choose the preferences that make the most sense to you regarding the controls' \textbf{predictability}, \textbf{efficiency}, or \textbf{explorability}:
    \begin{itemize}[label=$\circ$]
        \item \textbf{Predictability}: Allows you to obtain results with no surprises.
        \item \textbf{Efficiency}: Allows you to perform tasks with a minimum amount of time and effort.
        \item \textbf{Explorability}: Allows you to explore multiple possibilities to perform the task.
    \end{itemize}

    \item When making your selection, please consider the following \textbf{comprehensive factors}:
    \begin{itemize}[label=$\circ$]
        \item \textbf{UI control types}: Whether the UI control types make sense to you regarding predictability, efficiency, or explorability.
        \item \textbf{Scores}: How much each control is recommended in terms of predictability, efficiency, or explorability.
        \begin{itemize}[label=--]
            \item Each UI control has a score out of 10.
            \item In each option, a UI control with higher scores indicates a stronger recommendation.
            \item The total score of all UI controls in each option adds up to 10.
        \end{itemize}
        \item \textbf{Reasons}: Why each UI control is a good choice to allow predictability, efficiency, or explorability.
    \end{itemize}

    \item \textbf{Do NOT} base your preference selection on:
    \begin{itemize}[label=$\circ$]
        \item The number of UI control types (e.g., choosing an option simply because it offers more types of UI controls).
        \item The length of reasoning (e.g., choosing an option simply because the reasons are longer).
    \end{itemize}

    \item Remember, your preference selection should consider UI control types, scores, and reasons \textbf{collectively}. Feel free go back and interact with the UI controls if you want.
    \begin{itemize}[label=$\circ$]
        \item If UI control types are the same, consider the scores and reasons.
        \item If UI control types and scores are the same, consider the reasons.
    \end{itemize}
\end{itemize}

\end{document}